\documentclass[sigconf]{acmart}

\usepackage[linesnumbered,ruled,vlined]{algorithm2e}
\usepackage{amsmath}
\usepackage{graphicx}
\usepackage{subcaption}
\usepackage{csquotes}
\usepackage{hyperref}
\usepackage{arydshln}
\usepackage{multirow}

\newcommand{\specialhline}[1]{\noalign{\global\arrayrulewidth=#1}\hline}
\AtBeginDocument{%
  \providecommand\BibTeX{{%
    \normalfont B\kern-0.5em{\scshape i\kern-0.25em b}\kern-0.8em\TeX}}}

\copyrightyear{2025}
\acmYear{2025}
\setcopyright{cc}
\setcctype{by-nc}
\acmConference[DIS '25]{Designing Interactive Systems Conference}{July 5--9, 2025}{Funchal, Portugal}
\acmBooktitle{Designing Interactive Systems Conference (DIS '25), July 5--9, 2025, Funchal, Portugal}
\acmDOI{10.1145/3715336.3735824}
\acmISBN{979-8-4007-1485-6/2025/07}

\begin{document}


\title[DFX]{DesignFromX: Empowering Consumer-Driven Design Space Exploration through Feature Composition of Referenced Products}

\author{Runlin Duan}
\orcid{0000-0001-8256-6419}
\authornote{Both authors contributed equally to this research.}
\affiliation{%
  \institution{School of Mechanical Engineering \\ Purdue University}
  \city{West Lafayette}
  \country{USA}}
\email{duan92@purdue.edu}

\author{Chenfei Zhu}
\orcid{0009-0003-3408-2876}
\authornotemark[1]
\affiliation{%
  \institution{School of Mechanical Engineering \\ Purdue University}
  \city{West Lafayette}
  \country{USA}}
\email{zhu1237@purdue.edu}

\author{Yuzhao Chen}
\orcid{0009-0005-6196-1176}
\affiliation{%
  \institution{Elmore Family School of Electrical and Computer Engineering \\ Purdue University}
  \streetaddress{610 Purdue Mall}
  \city{West Lafayette}
  \state{IN}
  \country{USA}
  \postcode{47907}
}
\email{chen4863@purdue.edu}

\author{Yichen Hu}
\orcid{0009-0001-9385-4456}
\affiliation{%
  \institution{Department of Computer Science}
  \city{West Lafayette}
  \country{USA}}
\email{hu925@purdue.edu}

\author{Jingyu Shi}
\orcid{0000000151592235}
\affiliation{%
  \institution{Elmore Family School of Electrical and Computer Engineering \\ Purdue University}
  \city{West Lafayette}
  \country{USA}}
\email{shi537@purdue.edu}

\author{Karthik Ramani}
\orcid{0000-0001-8639-5135}
\affiliation{%
  \institution{School of Mechanical Engineering \\ Purdue University}
  \city{West Lafayette}
  \country{USA}}
\email{ramani@purdue.edu}


\begin{abstract}
Industrial products are designed to satisfy the needs of consumers. The rise of generative artificial intelligence (GenAI) enables consumers to easily modify a product by prompting a generative model, opening up opportunities to incorporate consumers in exploring the product design space. 
However, consumers often struggle to articulate their preferred product features due to their unfamiliarity with terminology and their limited understanding of the structure of product features.
We present DesignFromX, a system that empowers consumer-driven design space exploration by helping consumers to design a product based on their preferences.
Leveraging an effective GenAI-based framework, the system allows users to easily identify design features from product images and compose those features to generate conceptual images and 3D models of a new product.
A user study with 24 participants demonstrates that DesignFromX lowers the barriers and frustration for consumer-driven design space explorations by enhancing both engagement and enjoyment for the participants.

\end{abstract}



\begin{teaserfigure}
  \includegraphics[width=\textwidth]{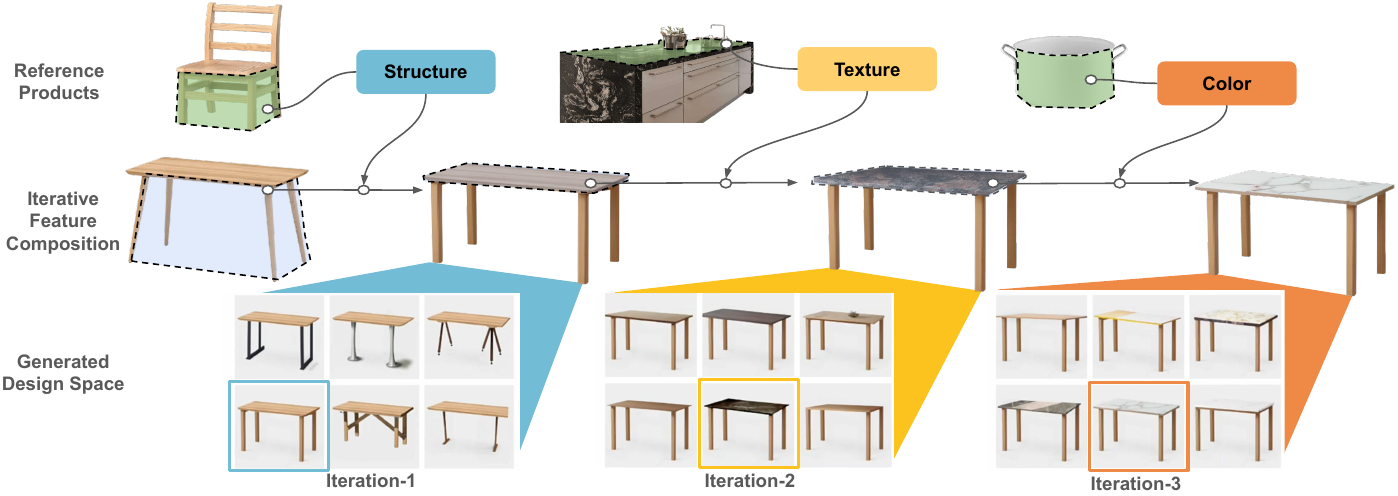}
  \Description[Design space exploration of a desk design using DesignFromX.]{This figure demonstrates the workflow of exploring desk designs using  DesignFromX. Initially, users select components from reference products—for example, taking the structural legs from a wooden chair. The system then integrates these chosen features into a new desk design. Over multiple iterations, users continue selecting additional features, such as texture from a countertop and color from a mug, progressively expanding and refining their personalized desk designs. Each iteration generates various design options, visually illustrating the evolving design space based on user selections.}
  \caption{
  Exploring the design space of a desk using DesignFromX. The process begins with the user selecting a component from a reference product image—here, the legs of a wooden chair. The system identifies and suggests design features of the selected component. The user then composes this feature, in this case, the structural form, into a designated part of the desk. Based on the user-defined composition, a Generative AI model generates a new design space for the desk. In subsequent iterations, the user can incorporate additional design features from other reference products to further explore the design space while retaining the features of their previous selections.
  }
  \label{fig:teaser}
\end{teaserfigure}


\maketitle

\section{Introduction}

Designing successful industrial products requires understanding the needs of consumers deeply and transforming their preferences into tangible product features \cite{jin2016understanding,jindal2016designed}, which are often represented in images, sketches, or 3D models.
Traditionally, exploring the design space of a product is tedious and time-consuming, involving professional designers proposing a wide range of potential features and iteratively composing them into a new product \cite{goel2005design,yang2009observations,piya2017co}.
Generative artificial intelligence (GenAI) introduces an efficient alternative by empowering consumers to actively participate in the design space exploration of industrial products. By generating images from prompts \cite{openai2024dalle3}, GenAI tools enable consumers to modify designs of products such as chairs, lamps, and bicycles by altering features such as style, color, texture, or even mechanical mechanisms.
However, many consumers face challenges in articulating and systematically organizing their desired features due to a lack of familiarity with design terminology and limited knowledge of the structural composition of industrial products.

References such as images, keywords, and documents have long been employed to convey desired design features and guide design composition \cite{chen2020qcue,
bae2020spinneret,chen2019collaborative}.
Typically derived from patent datasets \cite{sarica2020technet}, crowdsourced repositories \cite{shi2017data,dong2014knowledge,kim2012causality}, and organized open documents of existing products \cite{wang2023embedding,luo2019computer}, these references often showcase engineering details.
Although this emphasis on professional knowledge benefits expert designers who are familiar with design concepts and specialized terminologies, it makes such references less accessible to inexperienced designers, especially consumers seeking to influence the design process.

GenAI has been employed to generate high-quality references \cite{davis2024fashioning,chung2023promptpaint,hou2024c2ideas,wang2024roomdreaming} to help human designers in various design task.
Previous research on product design has leveraged GenAI to create references for 3D modeling \cite{liu20233dall}, prototyping \cite{zhang2024protodreamer}, generate early-stage concepts \cite{zhu2023generative,choi2024creativeconnect}, and decision making \cite{khan2023does,wang2023task}. 
However, current GenAI-based product design predominantly relies on keywords to prompt GenAI, posing challenges in generating product designs that incorporate structurally composed design features. 
For instance, the keyword "wheel" may represent a wide range of characteristics, including shape, color, structure, and mechanism. 
Combining these features with elements from other products can result in entirely distinct designs, underscoring the need for more intuitive and flexible systems.

We propose a novel design support system that empowers consumers to actively participate in shaping industrial products. To inform our design, we conducted a formative study (N=8) to observe how individuals with no design experience approach designing an industrial product. Their practices were monitored and discussed by industrial design experts, which helped us identify key design goals for our system. Drawing from the study feedback and discussions, we derived a set of design goals to develop a consumer-driven design support system that enables product consumers to explore the design space of their desired products. Here, the end-users of our system are considered as novice designer, refering to product consumers without prior knowledge or experience in industrial product design. To simplify the design process and encourage consumer engagement, our system focuses on early-stage design, where consumer participation has the greatest impact. 
Building on this foundation, we present DesignFromX, a consumer-driven design support tool powered by GenAI. 
The system is designed to (1) segment and analyze design features from reference product images, (2) support user-guided design composition by integrating design features from different products, and (3) provide an effortless experience on design space exploration for consumers to shape the desired products.

We conducted a within-subject user study (N=24) to evaluate the effectiveness and usability of our system in supporting consumers composing design features to their desired products.
The results showed that participants preferred DesignFromX over a baseline GenAI system due to its ability to provide a more effortless and engaging experience in exploring the design space. 
Quantitative data further indicated that DesignFromX helps consumers utilize broader categories of features and encourages exploration of a greater variety of new designs. 
Additionally, an evaluation conducted by expert designers highlighted that DesignFromX effectively captures consumer preferences when consumers are given the opportunity to design products themselves.

We list our contributions as follows:

\begin{itemize}
    \item A workflow of integrated machine learning models to segment and analyze design features from images of referenced products.
\end{itemize}
\begin{itemize}
    \item A GenAI-powered design support system that empowers consumers to explore the design space through feature composition of referenced products.
\end{itemize}
\begin{itemize}
    \item A user study validates the effectiveness of the system functions and the system usability.
\end{itemize}

\section{Related Work}

\subsection{Visual and Textual Reference for Design Space Exploration}

For decades, visual and textual references, also known as visual and textual cues \cite{chen2020qcue}, have been adopted to provide intuitive and organized design analogies that promote creativity and efficiency in exploring the design space of conceptual designs \cite{yang2009observations}.
Early works used language datasets to provide technology terms and keywords as textual references \cite{miller1995wordnet,fabian2007yago,liu2004conceptnet, speer2017conceptnet, kim2012causality,dong2014knowledge}.
Those datasets were grounded in daily common sense and failed to provide detailed product information in textual cues.
To support product design, the researcher built professional design knowledge datasets to provide more precise and comprehensive references \cite{qian1996function,kim2012causality,sarica2020technet}.

Grounded in those datasets, researchers were able to provide a combination of visual and textual references for designers, including mind maps \cite{chen2019collaborative, chen2019mini, chen2020qcue,bae2020spinneret} and design knowledge graphs \cite{han2022semantic,shi2017data,luo2018design,sarica2021design} constructed by semantic networks. 
Early visual references aimed to complement textual cues in visualizing the connection between designs in graphs.
Later works seek to provide design images to provide more direct details.
Wang et al.\cite{wang2023embedding} introduced a visual knowledge graph that provides images of experimental prototypes for reference. 
Other works use images references to promote human designers, including InnoGPS \cite{luo2019computer} and VISON \cite{song2022design}.
Also, image references have been widely adopted to promote human creativity in many design tasks, including iterative visual design \cite{chilton2021visifit}, early-stage ideation \cite{koch2019may,kazi2017dreamsketch}, fashion style exploration \cite{jeon2021fashionq}, and collaborative ideation \cite{lin2020your}. 

However, the previous visual and textual references are mostly retrieved from a pre-defined dataset grounded in existing knowledge and products, thus constraining the designer to precedents and raising a high demand for designer expertise and experience to reuse them.

\subsection{GenAI-based Design Support Tools}

Generative AI (GenAI) supports unconstrained design exploration and lowers the barrier for analysis and expression by reasoning with cross-domain knowledge and generating multi-modal representations, such as text and images \cite{rombach2022high,vaswani2017attention}.
Such advances in GenAI have inspired researchers to develop novel design support tools to increase creativity and prevent design fixations \cite{wadinambiarachchi2024effects}.
One challenge humans encountered was identifying a suitable text prompt to initiate the generation \cite{gao2020making,li2021prefix,mahdavi2024ai,wang2024promptcharm}. 
Many works focused on finding better prompts to generate visual illustrations, including experiments on text-to-visual prompts \cite{liu2022opal}, promoting support \cite{brade2023promptify}, promoting collaboration for image generation, editing, and sharing \cite{verheijden2023collaborative}, extracting keywords for visual generation \cite{son2024genquery, choi2024creativeconnect}.
Beyond using text prompting to generate visual illustrations, researchers have explored painting interactions \cite{chung2023promptpaint}, word-color associations \cite{hou2024c2ideas}, and embedding human preferences \cite{wang2024roomdreaming} to empower further human control over the visual design process. 
Davis et al. further studied controlling aesthetic attributes to facilitate generative visual illustration in fashion design \cite{davis2024fashioning} and self-expressions \cite{shi2024personalizing}.
Besides the atheistic consideration, recent research also leverages generative visual design to promote information sharing and understanding, such as designing infographics \cite{zhou2024epigraphics} and data analogies \cite{chen2024beyond}.

The advance of GenAI also facilitates the development of novel design support tools for physical product design.
Liu et al. first introduced 3Dalle \cite{liu20233dall}, a GenAI system that supports 3D modeling by generating visual references. 
Then, Lee et al. further investigate GenAI-assisted 3D modeling by comparing the sketch-based prompts and text-based prompts \cite{lee2024impact}.
Besides supporting 3D modeling for physical products \cite{urban2021designing}, GenAI also helped human designers in early-stage design, such as design generation \cite{zhu2023generative}, identifying key functions of a product \cite{wang2023task}, or design space explorations \cite{khan2023does}.

The previous works illustrate that GenAIs can promote generation, expression, and analysis in many creative tasks.
However, how to steer GenAI in supporting consumers, also identified as unexperienced designers in product design, remains unexplored. More precisely, how to help consumers acquire useful features from reference products and compose those features into their desired product.
In this paper, we aim to overcome the problem by leveraging GenAI to uncover the design features and then empowering an intuitive design space exploration framework using visual and textual references recommended by the GenAI-based system.

\subsection{Zero-shot prompting on Large Language Model}

Large language models (LLMs) \cite{bommasani2021opportunities,brown2020language,thoppilan2022lamda} have demonstrated extraordinary potential in enhancing human performance across a wide range of tasks, from learning \cite{cai2024advancing} and creative work \cite{petridis2023anglekindling,de2024llmr} to organizing and managing information \cite{jo2024understanding}. 
Zero-shot prompting \cite{betz2021thinking} allows users to describe tasks directly in natural language and receive answers from LLMs without additional training. 
In the field of human-computer interaction (HCI), researchers have leveraged zero-shot learning to support atomic tasks within human-AI collaboration scenarios. 
For instance, Chen et al. employed LLMs to recognize objects in tutorial videos \cite{chen2024tutoai}, while other works have integrated LLMs for object recognition using gestures and eye gaze \cite{lee2024gazepointar}.
Additionally, LLMs have been used in group discussion contexts, such as answering questions in classrooms \cite{liu2024classmeta} and evaluating ideas in group brainstorming sessions \cite{shaer2024ai}. 
To extend LLM capabilities for handling more complex, multi-step tasks, researchers have introduced the design of LLM chains, which scale prompts for each step of a procedural task \cite{wu2022promptchainer,wu2022ai}.

DesignFromX utilizes zero-shot prompting to recommend relevant design features for consumers. 
Beyond simply suggesting design features, the system prompts the LLM to identify components and generate descriptive designs. 
Similar to previous work where LLMs were used for atomic tasks such as summarization \cite{shang2023luse} and moment detection \cite{croitoru2023moment}, DesignFromX uses generated descriptions as input for other models within the design pipeline. 
To address potential hallucinations from the LLM, we incorporated human designers at each step, ensuring accuracy and relevance \cite{zhang2024siren}. 
Consistent with prior research, we validated the effectiveness of LLM outputs through human-in-the-loop evaluations.

\section{Design Rationale}\label{sec:design_rationale}

We developed the design rationales of DesignfromX based on a formative study aimed at understanding consumer needs and behaviors in product design. 
As the consumers are not typically the target users of design support systems, their performance and challenges are typically different from the expert designers \cite{chen2022behaviors,popovic2000expert}
By observing and interviewing participants with no prior design experience, we identified key patterns in how consumers approach ideation, visualization, and refinement of their designs. 
These insights guided the development of design goals that informed the creation of our system. In the following sections, we elaborate on these insights and demonstrate how they shaped the system's features.

\subsection{Formative Study}

We conducted a formative study to investigate consumer needs and behaviors in product designs. This study involved 8 participants (5 male, 3 female) recruited through the university email list and online advertisements, all without experience in product design. 
Over two sessions—a 20-minute design practice followed by a 10-minute interview—participants showcased their designs for three daily products (e.g., a lamp, a mug, or a chair) with pencils and paper while being guided to use a visual generative AI tool (DALLE3) to visualize them, and shared their experiences and thoughts on this process. We summarize the key interview questions for the formative study in the appendix.

\subsection{Study Insights}

Based on observations and interviews conducted during the formative study, we identified the following key insights into consumer behaviors and needs in product design.

\subsubsection{Study Insight 1: Barriers to Decomposing Visual References}

Participants frequently expressed the needs for visual references to guide their ideas, favoring appealing designs for inspiration rather than starting from scratch—a process they found challenging. 

\begin{quote}
\emph{"Seeing pictures of existing designs really helps me figure out how all the features fit together and gives me ideas for what I could change or make my own. Starting with an image of a product is such a good way to get going—it’s way less scary. Like, I don’t really know how to design a bike from scratch, but I definitely have lots of thoughts about what I don’t like about the ones out there now..." (P2)} 
\end{quote}

However, unlike design experts who critically analyze visual references to decompose key features, enhance functionality, or explore innovative ideas \cite{yang2009observations,song2022design}, consumers struggled to discern individual features of their interests within these references and tended to adopt the entire design to reflect their personal tastes or needs. 
This approach reflects that \textbf{product consumers lack the capabilities in decomposing individual features from visual references}.

\subsubsection{Study Insight 2: Challenges in Articulating Aesthetic and Functional Design Preferences}

Participants frequently describe their design preferences as being shaped by either aesthetic appeal (e.g., color, texture, or shape) or functionality (e.g., practicality, ease of use), aligned with recent research on product design \cite{han2021exploration,hagtvedt2014consumer}. 
However, participants often struggled to articulate their desired design features with clarity and precision, a challenge that became especially evident when applying GenAI to generate images for new products. 

\begin{quote}
\emph{“I think one of the issues is confidence. When I was working on my desk, I really liked the color and shape of a door I saw in a reference, and I wanted to get them into my desk. But it took me a long time to figure out how to describe what I wanted in a way that felt professional or fit naturally into the (design) process...” (P5)}
\end{quote}

This sentiment reflects a common dilemma observed in the study: \textbf{while consumers can articulate general preferences, they often struggle to frame these preferences in a "design language"} that bridges the gap between abstract inspiration and actionable design specifications. 
Tools that help consumers articulate and formalize their preferences could empower them to engage more confidently in the design process and make more informed decisions.

\subsubsection{Study Insight 3: Facilitation of Exploration by Examples}

For many consumers, images of products generated by GenAI serve as intrinsic examples in exploring the design space, demonstrating how various features can be integrated into cohesive products. These examples inspire and inform participants’ own creations. 
As one participant noted, "The examples generated by AI made me think of ideas I hadn’t even considered before" (P4). 
The study highlights that GenAI not only enabled participants to express their ideas but also motivated them to venture beyond their comfort zones and gain a clearer understanding of design principles, including proportions, scale, and material representation—principles that are particularly valuable for individuals with limited design experience or training.
However, the study also identified limitations in participants' use of GenAI. Many struggled to create effective prompts that could guide the AI toward generating relevant designs. Additionally, \textbf{participants found it challenging to organize and manage the iterations of generated designs}, often wanting to revisit and refine earlier outputs but lacking a straightforward way to do so.
These challenges highlight the need for tools that support better prompt creation and iteration management to fully leverage the potential of GenAI in consumers-driven design workflows.

\subsubsection{Study Insight 4: Preference of Increment Design}

We observe that breaking designs into smaller, distinct components greatly reduces the complexity and intimidation of the design process. \textbf{Participants consistently favored making small, incremental adjustments to specific parts of a product rather than undertaking a complete redesign}. This preference underscores how segmenting a product into manageable components can make the process more approachable and user-friendly \cite{kim2012impact}. Additionally, participants express a clear need to integrate desired design features into previously generated images instead of generating entirely new ones.

\begin{quote}
\emph{"I like to build my stuff (design) step by step, making small changes each time and seeing how it looks after adding something new. It’s frustrating when it (GenAI) completely overhauls my entire design just to add a new feature. Like, one time I had a chair, and all I wanted was to add a pink cushion—but it ended up changing the whole chair to a different one with a pink cushion. I wish there was an easier way to just tweak the parts I’m unhappy with while leaving the rest untouched..." (P1)}.

\end{quote}

\subsection{Design Goals}

To address the challenges and needs identified in the formative study, we derived the following design goals for our system, aimed at empowering consumer-driven design by providing non-expert users with intuitive approaches and tools to actively participate in product desgin.

\begin{itemize}
    \item \textbf{Enable Decomposition of Visual References as Design Inspiration (DG1).} Facilitate the decomposition of product images into distinct, manageable components, allowing consumers to explore and adopt specific design features from visual references. This empowers consumers to initiate their design process with relevant inspirations and customize them to better align with their needs and preferences.
\end{itemize}

\begin{itemize}
    \item \textbf{Enhance Articulation of Aesthetic and Functional Preferences (DG2). }Provide tools that help consumers express aesthetic and functional features with clarity and precision. By offering professional, multi-modal descriptions of individual components, the system bridges the gap between abstract inspiration and actionable design specifications.
\end{itemize}

\begin{itemize}
    \item \textbf{Support Design Exploration Through Examples and Iteration (DG3).} Help consumers gain insights into design principles to build up confidence and satisfaction. Additionally, includes tools to manage and refine iterations, enabling users to organize their workflow and revisit earlier outputs easily.
\end{itemize}

\begin{itemize}
    \item \textbf{Promote Incremental and Focused Design (DG4).} Simplify the design process by enabling consumers to make small, incremental adjustments to specific product components rather than undertaking complete redesigns. This includes mechanisms to modify selected part without unintended changes to other parts of the product, giving consumers precise control over their designs.
\end{itemize}

These design goals set the foundation for the system features described in the next section. By addressing the identified user needs, the system integrates advanced GenAI capabilities, empowering consumers to engage in intuitive, iterative, and preference-driven product design.
\section{DesignFromX}

We propose \textbf{DesignFromX}, a novel design support system that empowers consumers to actively participate in product design by leveraging GenAI to analyze design features from visual reference and compose them into new designs.
DesignFromX consists of three core modules, addressing proposed design goals: 
\begin{itemize}
    \item A component query module to locate and segment preferred components from reference product images (DG1,4);
    \item A feature analysis module to analyze the design features of the queried components (DG2,4);
    \item A design composition module to iteratively compose and generate new designs based on user-selected design features (DG3).
\end{itemize} 
We integrated these modules into an intuitive design exploration workflow to ensure accessibility for consumers (DG4).
In this section, we introduce the system workflow and illustrate detailed design considerations for each module.

\subsection{System Workflow}

\begin{figure*}[h]
  \centering
  \includegraphics[width=\linewidth]{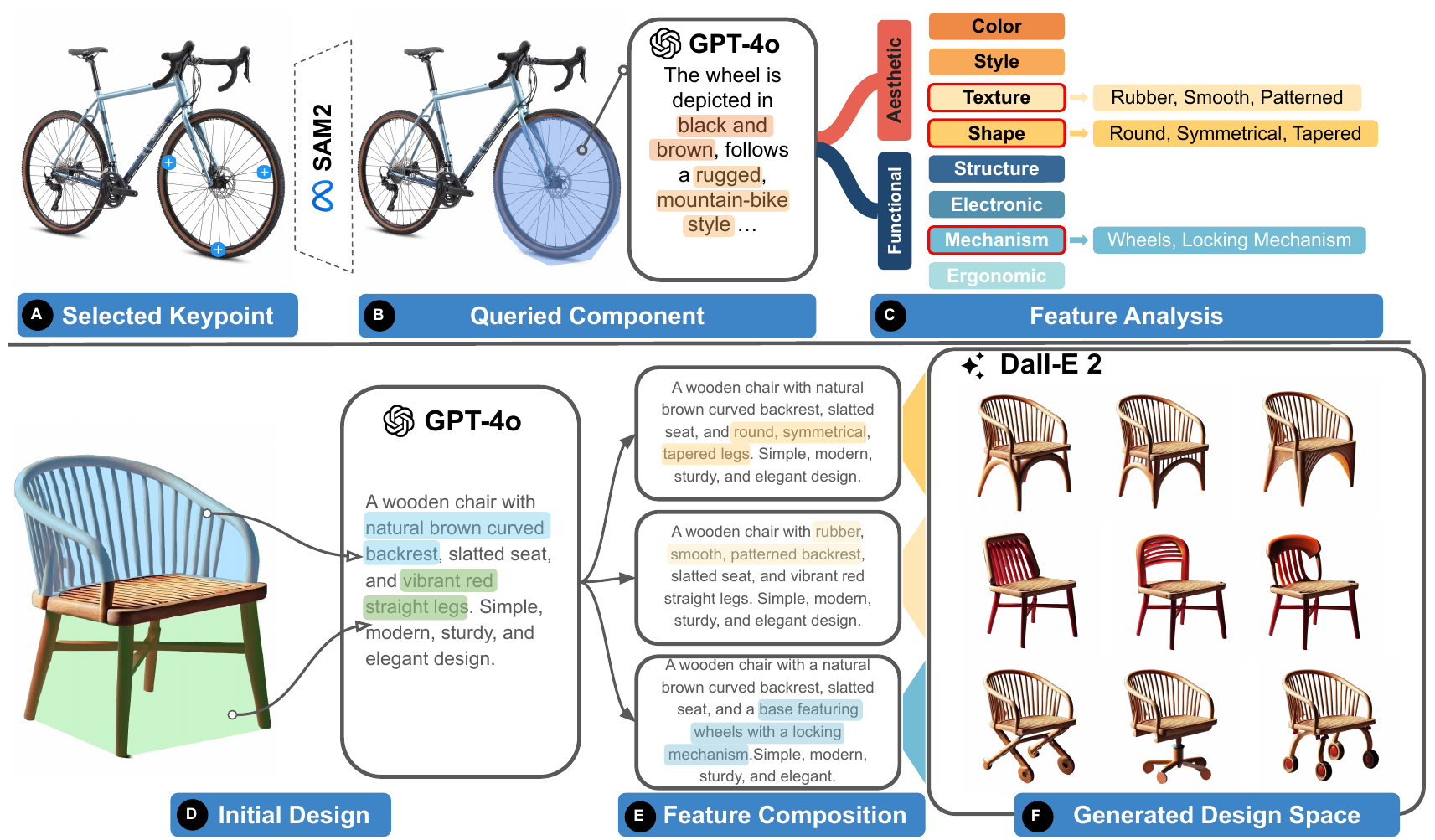}
  \Description[An overview of DesignFromX workflow.]{This figure provides an overview of the DesignFromX workflow through multiple steps. Initially, users select keypoints on reference images—for example, highlighting a bicycle wheel or parts of a chair (A). The system then segments and queries the chosen component (B), providing detailed feature analysis such as aesthetic properties (color, style, texture) and functional characteristics (structure, electronic, mechanisms, ergonomic features) (C). Users select specific analyzed features to integrate into their initial design (D). These features are composed into new design concepts (E), and finally, the system visualizes multiple generated design alternatives for users to explore (F).}
  \caption{An overview of DesignFromX workflow. (A) Users draw keypoints on the reference image to query a component; (B) The system segments and analyzes the queried component; (C) A feature analysis of the queried component is provided by the system, where users can choose the features they admire; (D) Users indicate the components or regions they want to modify on the initial design; (E) The system integrates selected design features into initial designs to compose new designs; (F) New designs are visualized for exploration.}
  \label{fig:workflow}
\end{figure*}

The workflow of DesignFromX, illustrated in Fig. \ref{fig:workflow}, supports consumers in exploring the design space of products and composing new designs in three key steps: 1) Reference Component Query; 2) Feature Analysis; and 3) Design Composition, based on an initial design and a reference image.

\textbf{Reference Component Query}: Users begin by clicking on the reference image to select key points on the desired component. The system then segments and analyzes the selected component, generating a detailed description. Fig. \ref{fig:workflow} (A)-(B) demonstrates this process, where the front wheel of a bicycle is selected as the queried reference component.

\textbf{Design Feature Analysis}: In this step, the system decomposes and analyzes the queried component, extracting its \textbf{aesthetic} (i.e., Color, Style, Texture, and Shape) and \textbf{functional} (i.e., Structure, Electronic, Mechanism, and Ergonomic) design features, as discussed in Section \ref{sec:design_rationale}. Users can explore these features and select the ones they find appealing. For example, in Fig. \ref{fig:workflow} (C), the user selects Texture, Shape, and Mechanism as the desired design features.

\textbf{Design Composition}: Once the desired features are selected, users determine how they should be incorporated into the initial design. They first use a brush tool to highlight the component or region they wish to modify. The selected features are then integrated into the textual description of the initial design to generate new design descriptions. Finally, the system visualizes the updated designs to showcase the expanded design space. Fig. \ref{fig:workflow} (D), (E), and (F) illustrate this process, where the selected texture, shape, and mechanism features of the bicycle wheel are incorporated into the legs and backrest of a chair.

By iterating these three steps, DesignFromX enables consumers to effectively explore the design space of and customize their products. In the following sub-sections, we outline the detailed design considerations for each module.

\subsection{Reference Component Query}\label{rcq}
The reference component query module comprises two core functions: (1) Interactive Image Segmentation and (2) Component Analysis.
For interactive image segmentation, we implemented Meta Segment Anything Model 2 (SAM2) \cite{sam2}, a state-of-the-art segmentation model known for its adaptability to unfamiliar objects. SAM2 offers exceptional interactive capabilities, enabling users to segment regions by simply clicking on the image, which facilitates real-time refinement of segmentation results. Our implementation adopts a click-based interaction method, where users can left-click to add areas and right-click to remove them, providing a seamless and intuitive segmentation experience.

For component analysis, we leverage a ChatGPT-4o \cite{openai2023gpt4} agent to analyze the segmented component. This function processes a splice of the original product image (e.g., Fig.~\ref{fig:workflow} (A)) alongside a copy where the queried component is masked with a transparent color overlay (e.g., Fig.~\ref{fig:workflow} (B)). The agent is prompted with the following request to analyze the masked component:
\textit{"This is a splice of two images. The left side shows an image of a product, and the right side is a copy where a region or component is covered by a transparent color mask. Analyze the component or the region masked."}

\subsection{Design Feature Analysis}

We propose a set of design features informed by previous studies and best practices in product design. These features not only enable a comprehensive analysis powered by generative AI (GenAI) but also provide consumers with a structured framework to better understand and explore various products. The proposed design features are categorized as follows:

\begin{itemize}
    \item \textit{Aesthetic features}: We considered the aesthetic design features as \textbf{\textit{shapes, colors, textures}}, following the taxonomy provided in \cite{crilly2004seeing}. 
    In addition, we involved \textbf{\textit{style}} as one of the design features, as it strongly influences how consumers perceive and feel the aesthetic of a product \cite{coates2003watches}.
    
    \item \textit{Functional features}: 

    The functional design features include \textbf{\textit{structure, mechanism, and electronics }}
    \cite{kim2012impact}. 
    Meanwhile, we especially involved \textbf{\textit{ergonomics}}, as it strongly impacts the user experience of products \cite{broberg1997integrating}.
\end{itemize}

We utilize a ChatGPT-4o agent to analyze the design features of the queried component. To facilitate this process, we provide the agent with a splice image consisting of two parts: one displaying the original product and the other showcasing the segmented component.
The agent is instructed to analyze the component within the context of the entire product, guided by the proposed design features.
We provide an example to standardize its responses and ensure consistency. 
Users then choose the design features they admire and compose new designs with those features in Design Composition.

\subsection{Design Composition}
To compose new designs incorporating reference design features, users first determine how these features should be integrated into their initial designs. This is facilitated through an intuitive brush tool available on the interface, which allows users to highlight the desired components or regions for modification.
We employ a ChatGPT-4o agent to first describe the initial design and identify the selected components, then to update the description by integrating the reference design features into the design of selected components or regions.
To be specific, the ChatGPT-4o agent is prompted to update the description with the following prompt: \textit{"Here is a description for a product: [Initial Design Description]. Update it based on the following changes: [Selected Components] is featured by [Selected Reference Features]"}.
The system then utilizes the editing function of DALL·E 2 model \cite{openai2024dalle3} to visualize the new designs by modifying the selected region of the initial design image based on the updated description (e.g., Fig.~\ref{fig:workflow} (E)).

\subsection{User Interface}
\begin{figure*}[h]
  \centering
  \includegraphics[width=\linewidth]{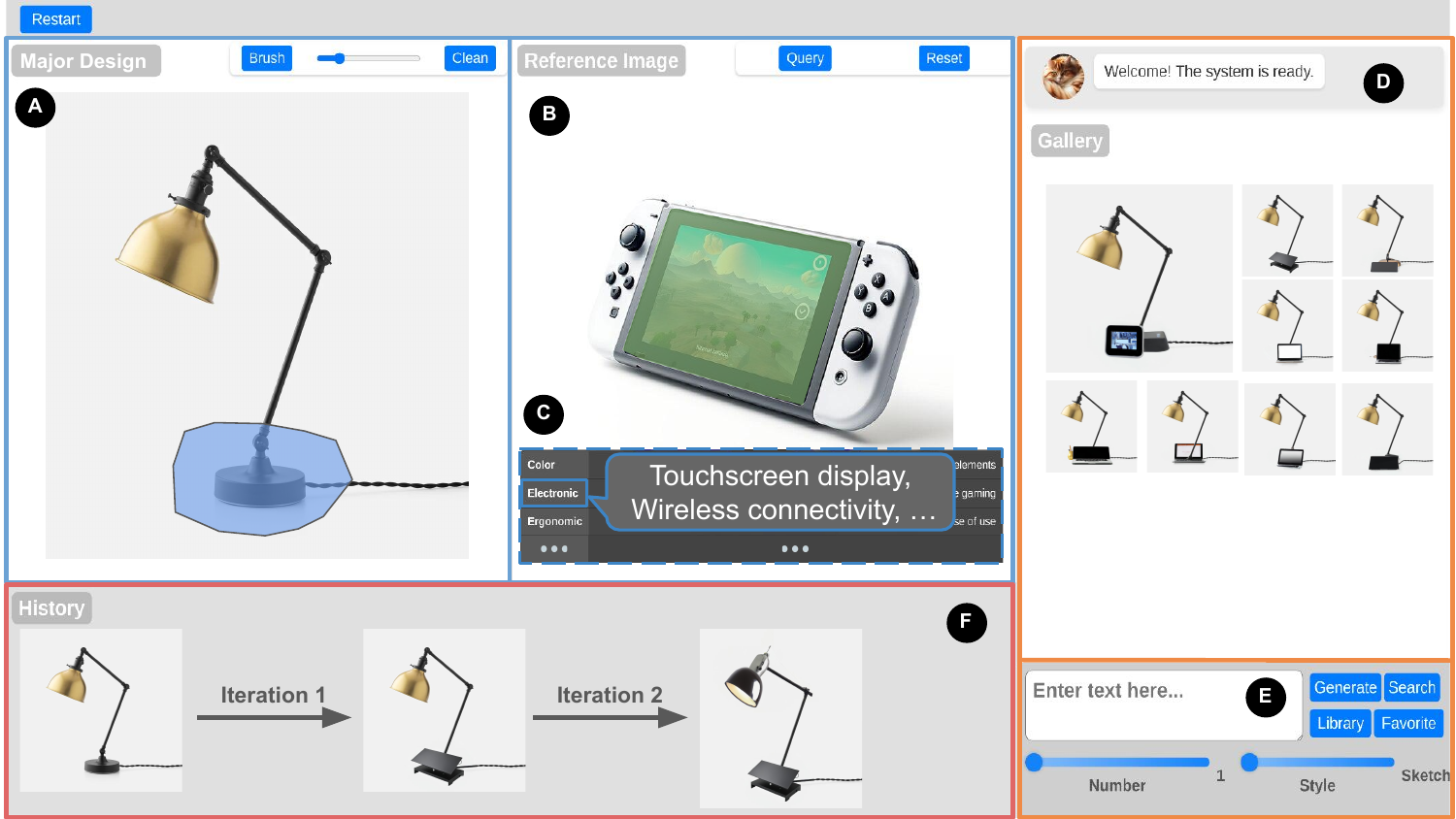}
  \Description[Illustration of DesignFromX interface.]{This figure presents the user interface of the DesignFromX system, highlighting its main components. Section A(left side) displays the primary design canvas with a lamp, allowing users to select and modify specific design components. Section B(top-center) shows a reference image of a handheld gaming device, which users query for design inspiration. Section C(below Section B) features an interactive pop-up describing specific design features, such as electronic components. Section D(top-right corner) includes an image gallery showcasing multiple generated variations of lamp designs. Section E(bottom-right corner) is a communication area where users input textual prompts to generate or search for additional designs. Finally, Section F(bottom-center to bottom-left) provides a design history, tracking iterative changes and allowing users to revert or modify past design steps.}
  \caption{DesignFromX system user interface. A) A major design canvas showcasing the updated design; B) A reference image canvas that enables interactions, like querying component, with reference images; C) an interactive pop-up design feature table, triggered by querying a component on the reference image; D) An image gallery to explore the design space of a product; E) A communication portal to search for reference images in various styles and from various sources; F) A history tracker that documents the changes made during design iterations, and allow for undo modifications.}
  \label{fig:ui}
\end{figure*}

DesignFromX system provides an intuitive HTML-based user interface to support consumers in product design, as shown in Fig. \ref{fig:ui}.
The interface consists of six functional regions: A) A major design canvas showcasing the updated design; B) A reference image canvas that enables interactions, like querying component, with reference images; C) an interactive pop-up design feature table, triggered by querying a component on the reference image; D) An image gallery to explore the design space of a product; E) A communication portal to search for reference images in various styles and from various sources; F) A history tracker that documents the changes made during design iterations, and allow for undo modifications.

\subsection{3D Visualization}

Our system supports generating the corresponding 3D model based on the product design formed in the previous steps, enabling users to visualize the design comprehensively and modify it intuitively.

A popular pipeline for 3D reconstruction from a single-view image is based on NeRF \cite{mildenhall2021nerf} and Gaussian Splatting \cite{kerbl20233d} techniques. These workflows reconstruct 3D models from images captured from different viewpoints, making their performance highly dependent on the quantity and quality of multi-view images. However, generating numerous images from a single input image may accumulate bias and increase inference time. Therefore, for each 2D product design, our system generates six images from different viewpoints, with each image produced by a 120-step multi-view diffusion model. Additionally, our system uses background-removed images as input to avoid artifacts and distortions.

\begin{figure*}[h]
  \centering
  \includegraphics[width=1\linewidth]{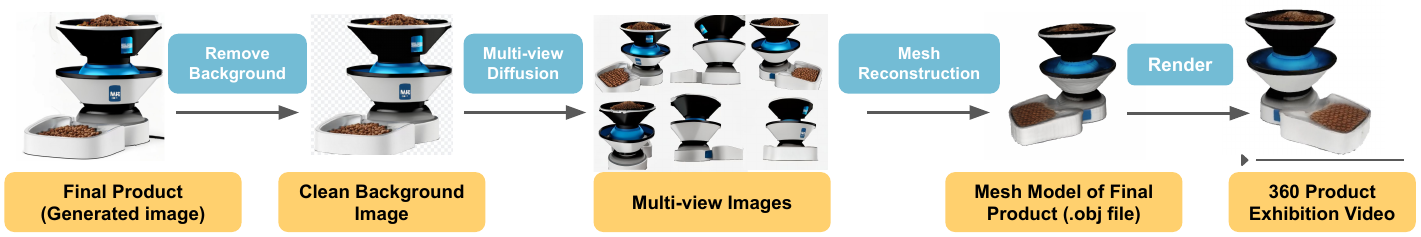}
    \Description[Workflow for converting a single image into a 3D model.]{This figure illustrates the workflow for generating a 3D model from a single product design image. The process starts with a generated product image, from which the background is removed to create a clean image. Next, multiple views of the product are produced using multi-view diffusion techniques. These multi-view images then undergo mesh reconstruction to create a detailed 3D model (in .obj file format). Finally, this 3D model is rendered into a 360-degree product exhibition video.}
  \caption{3D model generation from single design image}
  \label{fig: 3D_workflow}
\end{figure*}

Specifically, this extended module follows the workflow of InstantMesh \cite{xu2024instantmesh}, as illustrated in ~\autoref{fig: 3D_workflow}.

\subsection{System Implementation}
The implementation of DesignFromX is a web-based system with an HTML-based front-end user interface and a Flask-based back-end server. 
The whole system was run on a PC (Intel Core i9-10900KF CPU, 3.7GHz, 128 GB RAM, NVIDIA GeForce RTX 3090). 

\section{User Study}

We conducted a two-phase user study to evaluate the overall system effectiveness in supporting consumers in exploring the design space of a product. 
From this user study, we examined how the system fulfills our design goals: enabling the decomposition of useful design features from visual references (DG1), helping users articulate aesthetic and functional preferences (DG2), supporting the composition and iterative refinement of new design features (DG3), and promoting an engaging and intuitive experience for making incremental and focused adjustments to achieve the final design (DG4).
We especially pursued the answer to the following research questions:

\begin{itemize}
    \item \textit{RQ1: In what ways does DesignFromX enhance user interactions and performance during design space exploration?}

    \item \textit{RQ2: How does the user experience of interacting with DesignFromX differ from that of using the baseline system?}

    \item \textit{RQ3: How do the final products created with DesignFromX differ from those produced using the baseline system?} 
\end{itemize}

In this section, we describe the baseline system and then illustrate the study procedures, participants, and measures. We further discussed the result in the later section.

\subsection{Baseline system}

We deployed a baseline system that mirrored DesignFromX in its interface, AI models, and user interaction mechanisms without the fundamental functions of reference segmentation, feature analysis, and design composition. 
The same setup on the interface and models emphasized the comparison between the innovative functions of DesignFromX instead of the functionality of AI models.
Like DesignFromX, the baseline system allowed participants to search or use the generation tool to collect the reference images, ensuring consistency in the initial design stages.
With the baseline system, the user would manually analyze the design features in the reference images and describe the features to prompt the same GenAI model (Dalle-2) to generate images of the new design. 
Participants with the baseline system were provided with the same prompt templates that DesignFromX used for image generation as references.

In the next sections, we compared the results and the procedures using the baseline system and DesignFromX to seek the answer to the aforementioned research questions.

\subsection{Phase 1: Quality Assessment of System Modules}

In this phase, we assess the quality of the system modules in supporting the design space exploration of consumers. The system modules include the segmentation modules, the design feature analysis modules, and the composition modules. 
Each modules will be evaluated independently, with quantitative data supporting the results.
The primary goal of this evaluation is to validate whether the system features meet the intended design objectives and satisfy professional expectations.

\subsection{Phase 2: Comparing DesignFromX and the Baseline System}

\subsubsection{Participants} 

\begin{figure}[h]
  \centering
  \includegraphics[width=1\linewidth]{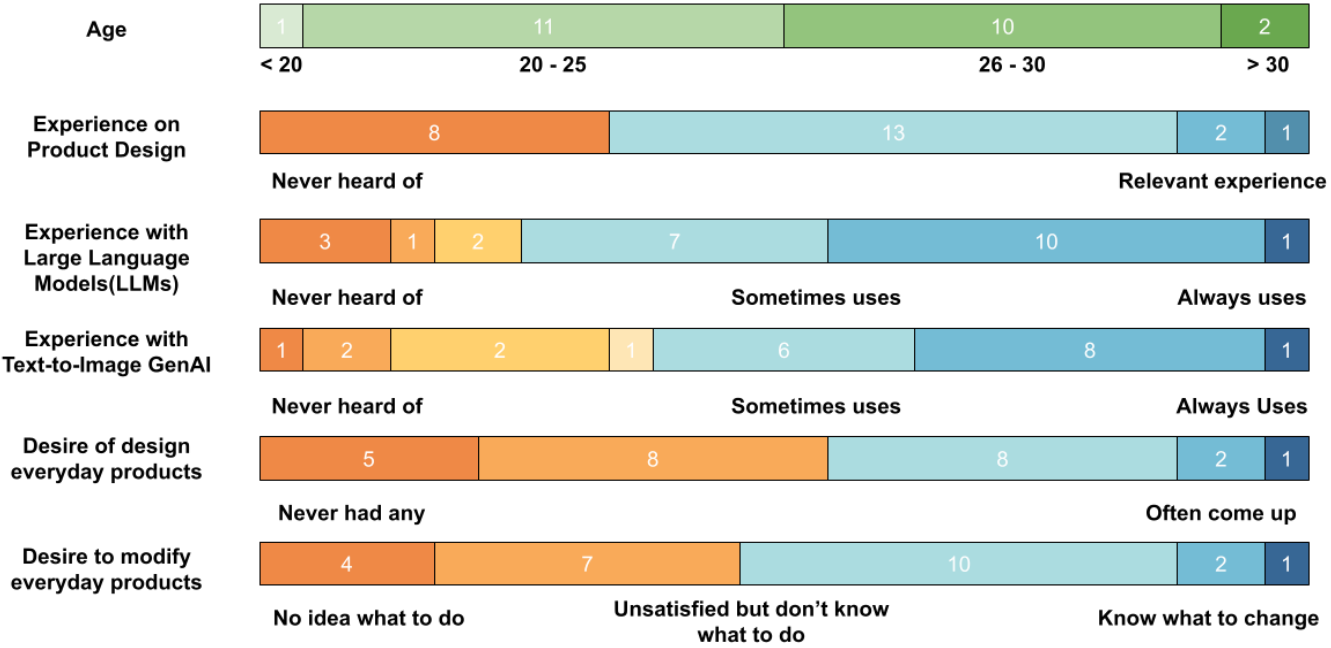}
  \Description[Participant demographics: age, design experience, AI usage, and design interests.]{This figure displays demographic information from participants in a user study. It presents data across several categories using horizontal bar charts. In the "Age" category, most participants are aged between 20 to 30. Regarding "Experience on Product Design," a majority have relevant experience. Most participants frequently use Large Language Models (LLMs) and Text-to-Image generative AI. In terms of "Desire for design everyday products" and "Desire to modify everyday products," most participants express sometimes wanting to design products but are uncertain about what to do with specific changes.}
  \caption{Demographics of the participants in the user study phase 2}
  \label{fig: demograghics}
\end{figure}

We recruited 24 participants (Male = 11, Female = 13) through online advertising and a university email list. 
To ensure the study's relevance to product consumers, we excluded individuals with formal education or work experience in mechanical engineering, industrial engineering, or related fields. 
nnnn period. Our use study is conducted with approval from our university's Institutional Review Board (IRB), \# IRB-2021-1159.
The detailed demographics of the participants are illustrated in ~\autoref{fig: demograghics}.

\subsubsection{Procedures} 

The second phase of the user study consisted of two sessions, where participants performed a product design task using two different systems: DesignFromX and a baseline system. The order of system usage was counterbalanced across participants to mitigate learning effects.

The study began with a brief interview (2 minutes) where participants were asked to reflect on daily products they disliked or wanted to customize, using the question: \textit{"Can you think of some daily products you don't like and want to customize?" }The participants then selected one of these products and either uploaded or searched for an image of the product to develop new designs.

Each participant used both systems in separate sessions, starting with a tutorial to familiarize themselves with the system's interface and features. Participants worked on the same design task across the two systems to ensure comparability.

After completing each session, participants filled out a system questionnaire to provide feedback on their experience. Finally, a post-session conversational interview was conducted to gather additional subjective feedback and insights into their interaction with the systems.

\begin{figure}[h]
  \centering
  \includegraphics[width=1\linewidth]{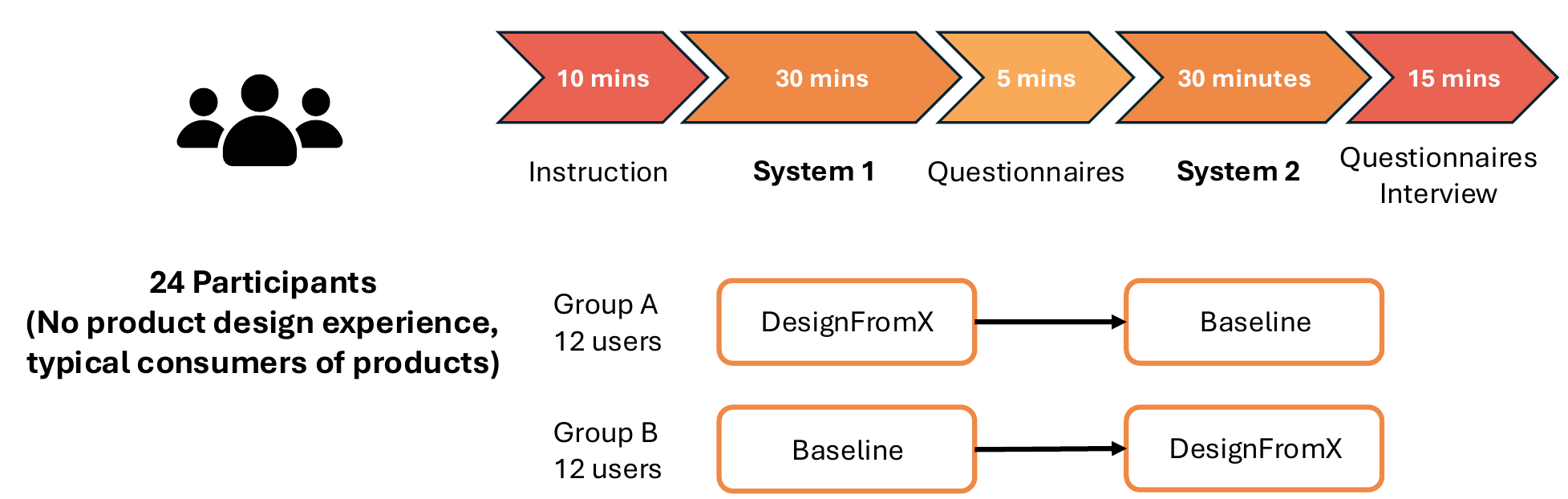}
   \Description[User study workflow comparing DesignFromX and Baseline systems.]{This figure depicts the detailed workflow for phase 2 of a user study involving 24 participants without prior product design experience. The workflow consists of several stages: first, participants receive 10 minutes of instruction, then they spend 30 minutes using the first system, followed by a 5-minute questionnaire. Afterward, they use a second system for another 30 minutes, and finally, they complete questionnaires and interviews for 15 minutes. Participants are divided into two groups of 12: Group A first uses "DesignFromX" then the "Baseline" system, while Group B first uses the "Baseline" system then switches to DesignFromX.}
  \caption{Workflow of the user study Phase 2}
  \label{fig: workflow_p2}
\end{figure}

The detailed workflow of the Phase 2 are illustrated in ~\autoref{fig: workflow_p2}.

\subsection{Measures}

We adopted quantitative and qualitative measures to evaluate the product design result, procedure, and user experience with DesignFromX and the baseline system.

\subsubsection{Quality of System Modules}

Evaluating the quality of system modules is essential to ensure their effectiveness and alignment with the overall design goals. 
To access the machine learning models for component segmentation and labeling, we employed two key metrics: the widely used Intersection over Union (IoU) and the accuracy of labeled segmented product components. The ground truth masks and labels were manually annotated by human designers, distinct from those involved in the comparative study. In total, 228 valid image masks were collected and labeled for the component segmentation evaluation.
For the design feature analysis and composition modules, we adopted human-reported metrics based on recent studies evaluating the quality of large language model responses \cite{lee2022evaluating}. These metrics, rated on a five-point scale, include accuracy, fluency, relevance, reporting correctness of the designated feature, readability of the generated text, and alignment with the given component.

\subsubsection{Evaluation of User Interaction and Performance}
To assess user performance in design space exploration, we introduced quantitative measures to compare the outcomes between the DesignFromX system and a baseline system. User log data was collected to analyze how participants interacted with the systems during their exploration process. Specifically, we measured the number of design features identified by participants, the number of new designs generated, and the number of major design iterations completed. In this context, a "major design iteration" refers to a significant modification of a particular design feature, indicating that the user is satisfied with the current state of the design and is ready to move forward with changes to the next feature. Additionally, we classified the design features identified by participants by analyzing the prompts used to generate images of new designs.

The final designs created by users were evaluated by experts using a 5-point Likert scale based on the following factors: 1) Feasibility, 2) Functionality, 3) Aesthetic, 4) Creativity, and 5) Overall quality. These evaluation criteria were derived from established research on product design assessment \cite{kudrowitz2013assessing,luchs2011perspective,sylcott2013understanding}. To complement the expert evaluations, we also collected self-evaluated scores from participants using the same set of factors, as suggested by prior studies on triangulating evaluation methods \cite{carroll2012triangulating}.

Detailed results and analysis on user interactions are provided in ~\autoref{sec:quantitative_results}, and the final product design quality in ~\autoref{sec:final_results}

\subsubsection{Evaluation of User Experience}

To compare the user experience between DesignFromX and the baseline system, we collected qualitative data through a post-session questionnaire. 
The qualitative metrics captured various aspects of user interaction and satisfaction with the systems.
As both systems incorporate a GenAI module into the design process, we assessed participants' self-perceived experience with the machine learning models \cite{wu2022ai}. 
In addition, we developed a Design Support Index, inspired by the Creativity Support Index \cite{cherry2014quantifying}.
This index focuses on evaluating participants' experiences from a novice support perspective, with metrics tailored to prioritize accessibility and engagement for users, rather than creativity enhancement.

To further compare the two systems, we employed task load metrics using NASA-TLX \cite{hart1988development} and usability metrics through the System Usability Scale (SUS), enabling a comprehensive evaluation of task load and usability.

The detailed findings on user experience evaluation are provided in ~\autoref{sec:user_exp}.

\subsubsection{Statistical Tests} 

We collected various paired data from DesignFromX and the baseline systems, including ranking data and several user operation logs. Meanwhile, we performed the Shapiro–Wilk test on all paired data and found that most of them had evidence to reject their normality. Therefore, in the following statistical test, we used the Wilcoxon signed-rank test, a commonly used non-Gaussian paired test, to determine whether there is any significant difference between the two systems and report their p-values and the effect size r-values computed using Rosenthal’s formulation. In this paper, the effect size r quantifies the score of DesignFromX with respect to the Baseline, where a positive r value indicates that DesignFromX has a higher value relative to the Baseline, and a negative r value indicates that DesignFromX has a lower value.

\section{Result}

\subsection{Supporting Design Composition}
\label{sec:quantitative_results}
We collected the user log information to understand the participants' activities during their design process.
From the data collected, we draw a discussion on the research question \textbf{\textit{RQ1: In what ways does DesignFromX enhance user interactions and performance during design space exploration?}} 

\subsubsection{Explore design features}

\begin{table*}[ht]
\begin{tabular}{c | c | c c | c c | c c c}
\specialhline{0.42pt}
\hline
\specialhline{0.4pt}
 \multicolumn{2}{c|}{} & \multicolumn{2}{c|}{DesignFromX} & \multicolumn{2}{c|}{Baseline} & \multicolumn{3}{c}{Statistics}\\
 \cline{3-9}
  \multicolumn{2}{c|}{}  &  Mean & Std & Mean &Std & p & r & Sig \\
 \hline
  \multicolumn{2}{c|}{\# of features explored} & 8.333 & 2.967 &6.625 & 2.927 & 0.023 & 0.425 & *\\
 \multicolumn{2}{c|}{ Types of features} & 4.917 & 1.037 & 4.417 & 1.152 & 0.088 & 0.391 &  \\
 \hline
\multirow{4}{*}{ \# of aesthetic features}   & Shape & 0.958 & 1.136 & 1.125 & 1.013 &  0.499 & -0.175& \\
                                    & Color & 1.333 & 1.027 & 1.500 & 1.041 & 0.552 &-0.154 & \\
                                    & Texture & 1.25 & 0.968 & 1.125 & 1.166 & 0.585 & 0.125 & \\
                                    & Style & 0.833 & 0.943 & 0.667 & 0.799 & 0.381& 0.234&  \\
\hline
\multirow{4}{*}{\# of functional features} & Mechanism & 1.458 & 1.19 & 0.708 & 0.841 & 0.038 &0.453 & *\\
                                    & Electronic & 0.542 & 0.706 & 0.292 & 0.538 & 0.109 & 0.483 & \\
                                    & Structure & 0.917 & 0.812 & 0.750 & 0.777 & 0.392 &0.247& \\
                                    & Ergonomic & 1.042 & 1.207 & 0.458 &  0.644 & 0.014 &0.739& * \\

\specialhline{0.42pt}
\hline
\specialhline{0.4pt}

\end{tabular}

\caption{Features attempted to compose using our and baseline systems. Meanwhile, the r‑value represents the effect size; \\ * indicates groups with significant differences (p<=0.05).}
\Description[Comparison of design feature exploration between DesignFromX and the Baseline System]{This table compares aesthetic features (color, style, texture, shape) and functional features (structure, ergonomic, mechanism, electronic) explored by participants using two systems: DesignFromX and a baseline system. Participants using DesignFromX explored significantly more features overall (average of 8.333) compared to the baseline (average of 6.625). They also explored more functional features, specifically mechanisms and ergonomics, with significant statistical differences (p≤0.05) and moderate to strong effect sizes (r-values of 0.453 and 0.739, respectively). However, there were no significant differences observed in the other features between the two systems.}
\label{tab: feature_to_compose_img}
\end{table*}

\begin{figure*}[h]
  \centering
  \includegraphics[width=\linewidth]{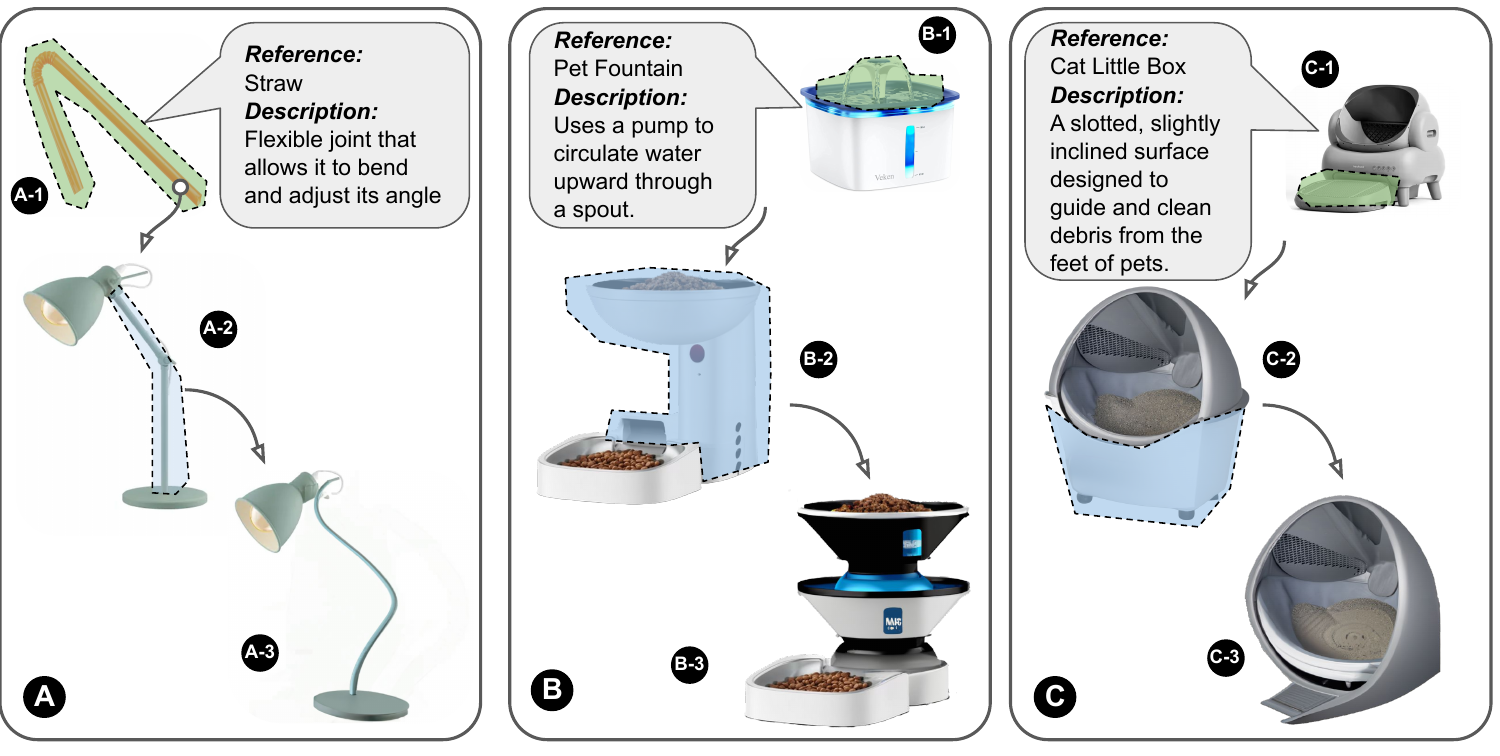}
  \Description[Examples of innovative product designs inspired by analyzing reference products.]{This figure illustrates three examples demonstrating how DesignFromX helps users integrate functional features from reference products into new designs: Example A shows a flexible desktop lamp inspired by a straw's bendable joint, allowing it to adjust angles easily. Example B depicts a dual-purpose pet feeder, combining features of a pet fountain (using a pump for water circulation) and a traditional feeder. Example C presents a cat litter box enhanced by adding a slotted, inclined ramp at the entrance, inspired by a reference litter box designed to clean pets' feet. Each example visually details how reference product features are analyzed, selected, and integrated into final innovative designs.}
  \caption{DesignFromX enables participants to compose more functional features to explore the design space, compared to the baseline system. (A) A participant analyzed the mechanism of a straw using DesignFromX and applied it to create a flexible stand for a desktop lamp. (B) Another participant designed a dual-purpose cat feeder by integrating a cat fountain with a traditional feeder. (C) A third participant enhanced a cat litter box by adding a ramp to the entrance, inspired by a similar product. }
  \label{fig:userstudy_goodexamples}
\end{figure*}

As shown in~\autoref{tab: feature_to_compose_img}, DesignFromX significantly increased the number of features explored by participants (DesignFromX: M=8.333, SD=2.967; Baseline: M=6.625, SD=2.927; p=0.023, r=0.425). 
We further categorized the design features explored by participants across both systems, and the results revealed that DesignFromX significantly improved the quantity of mechanism (DesignFromX: M=1.458, SD=1.19; Baseline: M=0.708, SD=0.841; p=0.038, r=0.453) and ergonomic (DesignFromX: M=1.042, SD= 1.207; Baseline: M=0.458, SD=0.644; p=0.014, r=0.739) features explored by participants among the eight feature categories. 
We demonstrate three examples of how the participants use DesignFromX to compose the mechanism features, as shown in ~\autoref{fig:userstudy_goodexamples}.
These improvements were because the system leveraged the LLM agent to analyze user-specified components and help identify relevant design features. 
One participant remarked, \emph{“In the beginning, I didn’t really know what I could do—I’m not an expert at analyzing features. But having an AI there really helps me think about what I can change and explore...” (P10).}
For the other features, the reason for the non-significant results was that these features were relatively easy for participants to describe independently. 
For example, people typically had little difficulty articulating design features such as the shape and color of a product, even without assistance from our system.

\begin{figure}[htp]
  \centering
    \includegraphics[width=1\linewidth]{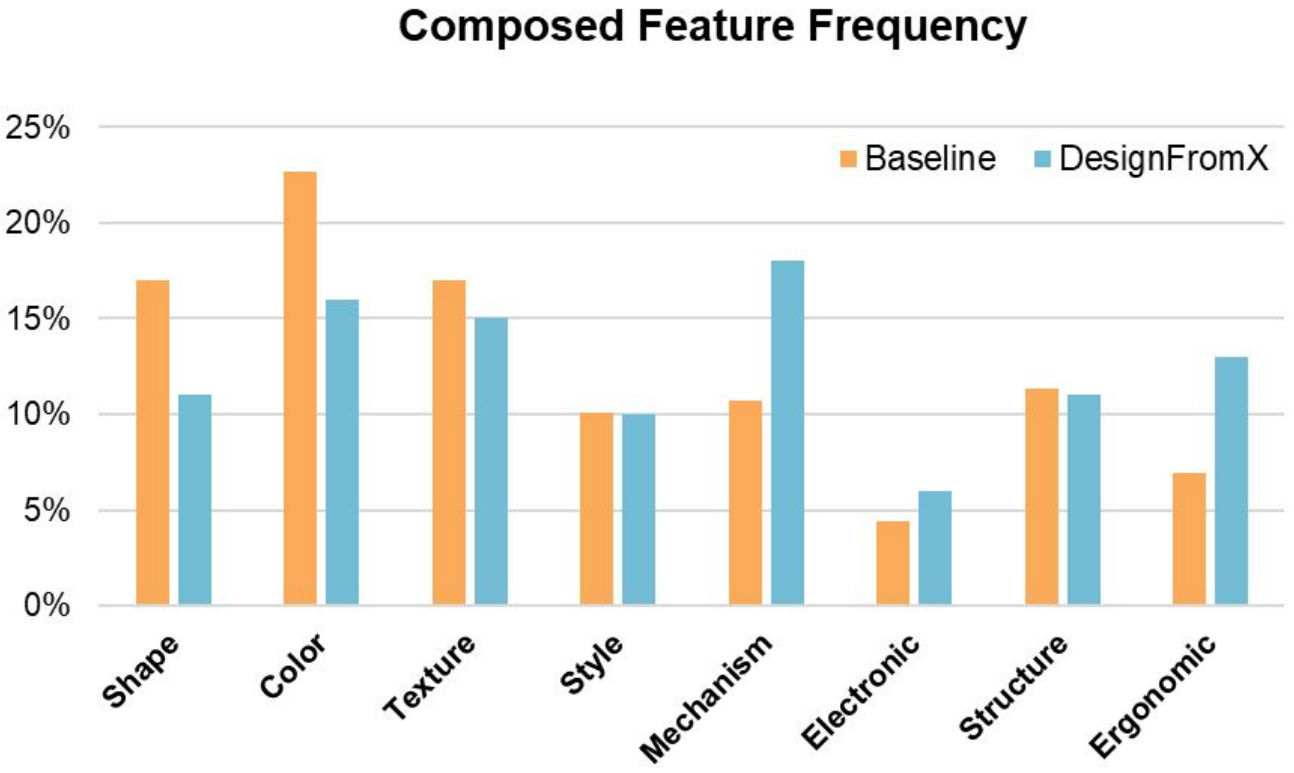}
    \Description[Comparison of composed design feature frequencies]{The figure is a bar chart comparing two systems, Baseline and DesignFromX, based on how frequently specific design features were composed by users. The horizontal axis lists eight design features: Shape, Color, Texture, Style, Mechanism, Electronic, Structure, and Ergonomic. The vertical axis represents frequency percentages, ranging from 0\% to 25\%. Overall, DesignFromX shows stronger user preference for composing Mechanism, Structure, and Ergonomic features, whereas the baseline system shows higher frequencies for composing Shape, Color, and slightly higher for Electronic.}
     \caption{The frequency of design features composed for our and the baseline system
     }
    \label{fig: pie_chart}
\end{figure}

Additionally, we compared participants' feature preferences between DesignFromX and the baseline system, as in ~\autoref{fig: pie_chart}. 
The comparison illustrated the distribution of focus across the eight types of design features. 
Overall, participants displayed a more balanced focus on functional and aesthetic features using DesignFromX than using the baseline system.
The reason for such a result was that DesignFromX helped designers to identify features that were difficult to describe.

\subsubsection{Supporting iterative design}

\begin{table*}[ht]
\begin{tabular}{c | c | c c | c c | c c c}

\specialhline{0.42pt}
\hline
\specialhline{0.4pt}
 \multicolumn{2}{c|}{}  & \multicolumn{2}{c|}{DesignFromX} & \multicolumn{2}{c|}{Baseline} & \multicolumn{3}{c}{Statistics}\\
 \cline{3-9}
  \multicolumn{2}{c|}{}                                     &  Mean     & Std       & Mean  & Std       & p        & r & Sig \\
 \hline
 & \multicolumn{1}{c|}{Time Spending(Minutes)}  & 25.694    & 11.842    & 20.885 & 6.192    &  0.056 &  0.389 & \\

 \hline
 \multirow{2}{*}{Major Design}  & \# of iterations          & 7.75     & 3.139     & 6.417 & 3.135     & 0.072 & 0.392  & \\

                                & Minutes per Iteration         & 3.353     & 0.999     & 3.711 & 1.433      & 0.455 & -0.152  & \\
\hline
\multirow{2}{*}{Design Composition}    & \# of new designs         & 81.667    & 27.789    & 66.25 & 29.271    & 0.042 & 0.425   & * \\
                             &  Minutes per Image Generated   & 0.314     & 0.102     & 0.343 & 0.098     & 0.229 & -0.245    & \\

\specialhline{0.42pt}
\hline
\specialhline{0.4pt}
\end{tabular}

\caption{Statistics of base information using our and baseline systems. Meanwhile, the r‑value represents the effect size;\\
* indicates groups with significant differences (p<=0.05).}
\Description[Comparison of user interaction metrics between DesignFromX and the baseline system]{This table compares user interactions between the DesignFromX system and a baseline system on several metrics: Participants spent slightly more time on average with DesignFromX (25.69 minutes) compared to the baseline (20.89 minutes), though this difference wasn't statistically significant.  In design composition: users generated significantly more new designs using DesignFromX (average 81.67 designs) compared to the baseline (average 66.25 designs), indicating a meaningful advantage, with a moderate effect size (r = 0.425, p = 0.042). For major design iterations, participants made slightly more iterations on average with DesignFromX (7.75) compared to baseline (6.42), but this wasn't statistically significant either. The average time spent per iteration or per generated image did not differ significantly between the two systems.}
\label{tab: basic_info_for_systems}
\end{table*}

Table \ref{tab: basic_info_for_systems} showed that participants generated significantly more new designs when using DesignFromX than the baseline system (DesignFromX: M=81.667, SD=27.789; Baseline: M=66.25, SD=29.271; p=0.042, r=0.425).
Here, each generated design referred to an image produced by the GenAI agent in response to the participant's request to compose new features into the current design.
During the study, we observed that DesignFromX reduced the time and efforts to identify and describe design features by the participant, empowered a seamless procedure in compositing new designs.
However, there was no significant difference in the number of iterations on the major design between the two systems. 
Here, "major design" referred to the design feature selected by the participants, indicating their satisfaction with the current design stage and their readiness to proceed to the next major changes.
The reason was that during the study, participants spent most of the explore the design space of each iteration.
They conducted several design compositions and generated many new designs as candidates before making a decision on the next major changes.
The standard deviation for both systems reflected a relatively wide variation in the number of new designs created, likely attributable to differences in participants' creativity and motivation.

\subsection{User Experience}
\label{sec:user_exp}

We collected qualitative results to compare the user experience of working with DesignFromX and the baseline system.
To answer the research question: \textbf{\textit{"RQ2: How does the user experience of interacting with DesignFromX differ from that of using the baseline system?"}}, we considered the metrics including the self-perceived experience in working with GenAI, the design support index using both tools, and the overall task workload.

\begin{figure}[htp]
    \centering
    \begin{subfigure}[b]{0.95\linewidth}
        \centering
        \includegraphics[width=\linewidth]{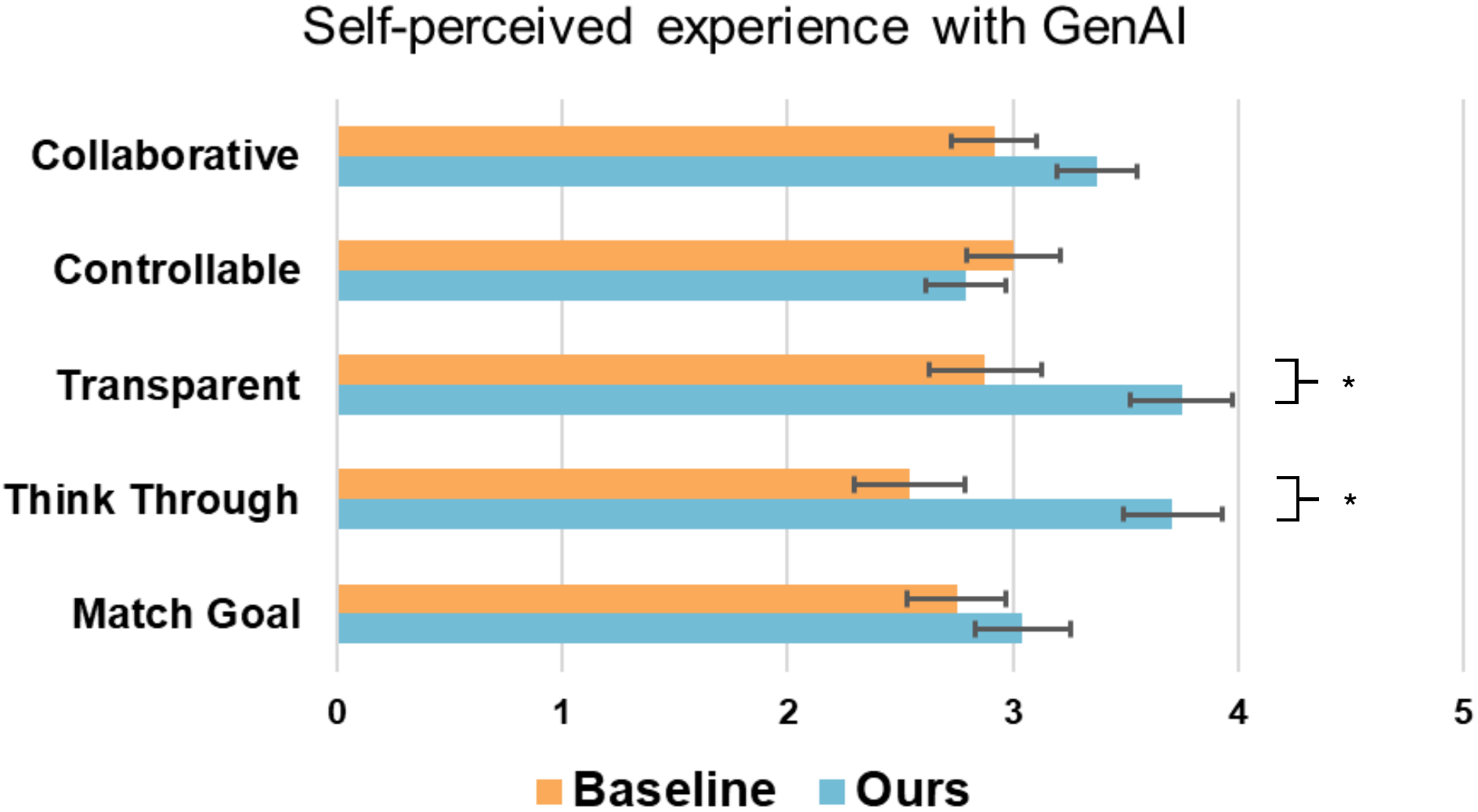}
        \caption{Statistics of self-perceived experience with GenAI using DesignFromX and baseline systems.
     }
        \label{fig:subfig2}
    \end{subfigure}
    \hfill
    \begin{subfigure}[b]{1\linewidth}
        \centering
        \includegraphics[width=\linewidth]{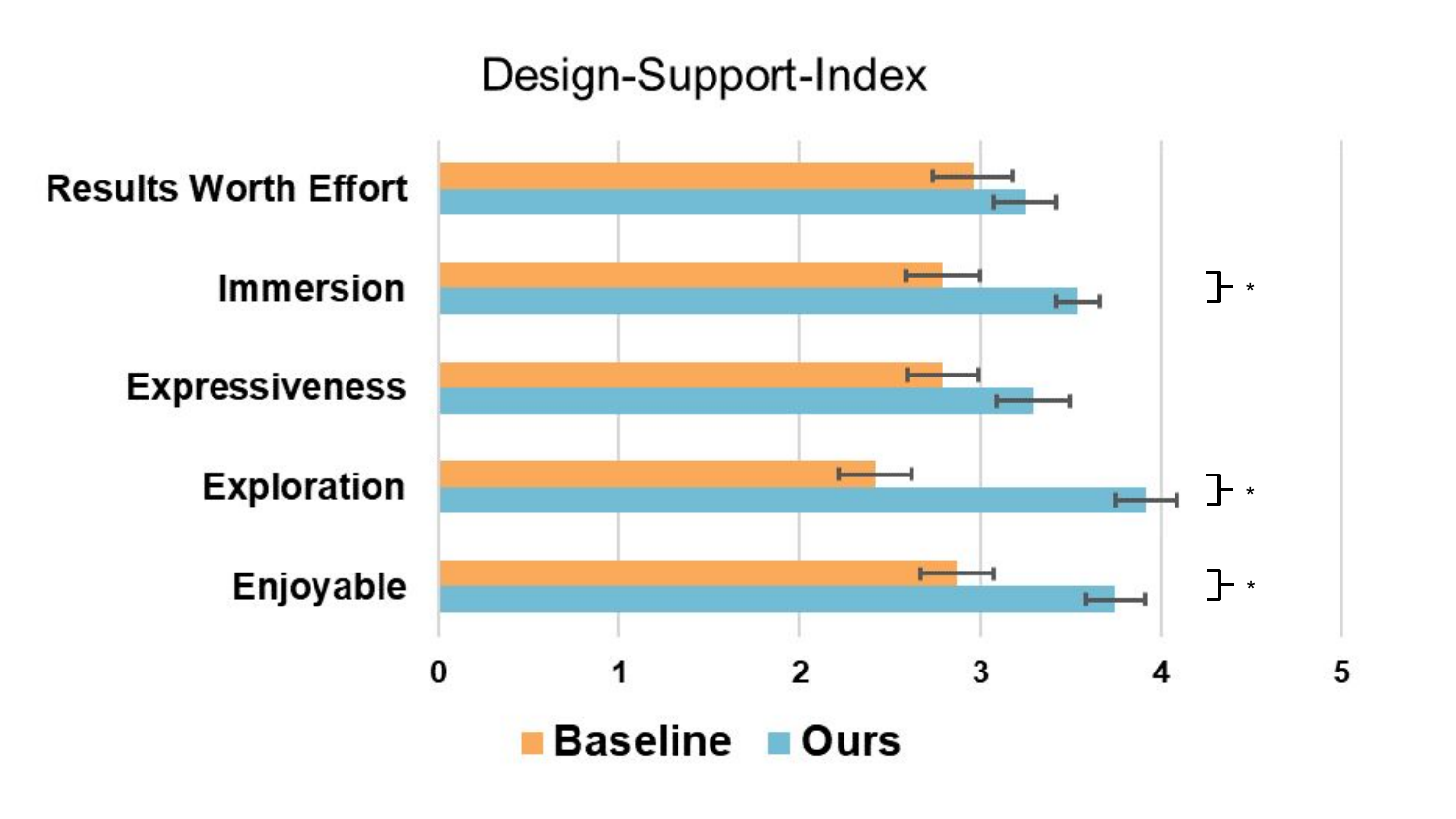}
        \caption{Statistics of design support index using DesignFromX and baseline systems.}
        \label{fig:designindex}
    \end{subfigure}
    \Description[Comparison of user experiences and design support.]{The subfigure (a) compares participants' self-perceived experiences using Generative AI (GenAI) between the DesignFromX system and the baseline system across five dimensions: Collaborative, Controllable, Transparent, Think Through, and Match Goal. Participants rated their experiences on a scale from 1 to 5, with higher scores indicating a more positive perception. The DesignFromX system significantly outperformed the baseline in the Transparent and Think Through categories. However, there were no significant differences observed in the Collaborative, Controllable, and Match Goal dimensions between the two systems. The subfigure (b) compares participants' ratings of the Design Support Index between the DesignFromX system and the baseline system across five factors: Results Worth Effort, Immersion, Expressiveness, Exploration, and Enjoyable. Participants rated these factors on a scale from 1 to 5, with higher scores being better. The DesignFromX system scored significantly higher than the baseline in four factors: Immersion, Expressiveness, Exploration, and Enjoyable, indicating it provided users with a richer, more engaging, and satisfying design experience. Results Worth Effort showed no significant difference between the two systems.}
    \caption{(a) Statistics of self-perceived experience with GenAI using DesignFromX and baseline systems. * indicates groups with significant differences (p<=0.05).\\
    (b) Statistics of design support index using DesignFromX and baseline systems. * indicates groups with significant differences (p<=0.05).}
    \label{fig:self_exp_genai}
\end{figure}

\subsubsection{Collaborating with GenAI agents}

As shown in Figure \ref{fig:self_exp_genai}, participants reported a significant improvement in the transparency of the system’s suggestions and decisions when using DesignFromX (M=3.750, SD=1.113) compared to the baseline system (M=2.875, SD=1.227; p=0.013, r=0.502). 
This improvement in transparency was attributed to the organized visualization of feature analysis results, which enhanced participants' understanding of the references and the LLM-generated suggestions. 
One participant noted, \textit{“Seeing a list of design features helps me understand exactly what I can ask the GenAI to modify.” (P2)}.
Additionally, the results demonstrated a significant improvement in guiding participants to think through and explore alternative designs (DesignFromX: M=3.708, SD=1.083; Baseline: M=2.542, SD=1.215; p=0.002, r=0.607).
This was because the system provided detailed descriptions of design features for the participant to consider their preferences on the new design.

DesignFromX helped participants structure their thinking and systematically explore different design compositions using the recommended design features.
In contrast, baseline users were observed spending more time formulating, often arbitrarily combining design features. 
During the study, some participants using the baseline system experienced design fixation, repeatedly focusing on a single feature instead of experimenting with others.

\begin{figure*}[h]
  \centering
  \includegraphics[width=\linewidth]{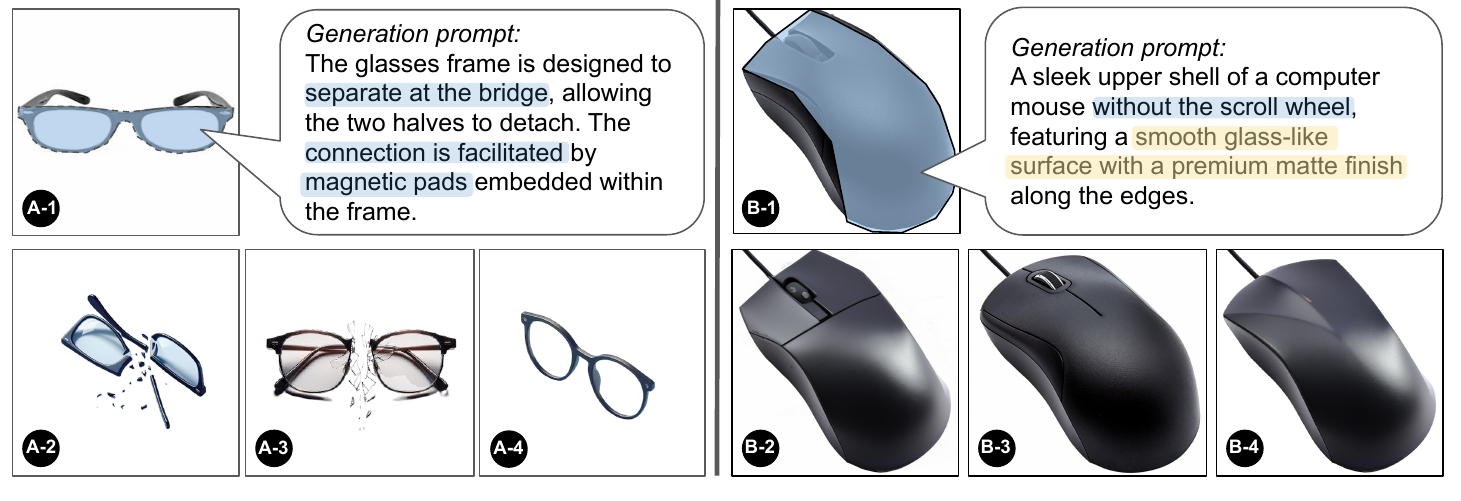}
  \Description[Failure examples in generation: glasses and mouse.]{This figure illustrates examples of unsuccessful attempts to generate specific design ideas using Generative AI (GenAI): Example A depicts a user's prompt for glasses with a detachable frame connected by magnetic pads. The images generated failed to accurately capture the intended design: they either incorrectly represented glasses that appear broken or separated unnaturally, rather than clearly showing a functional magnetic connection. Example B illustrates another participant's attempt to generate a computer mouse without a scroll wheel and with a smooth, matte finish. The resulting images incorrectly included the scroll wheel or did not fully match the described surface features. These examples highlight current limitations of GenAI in accurately interpreting and generating unconventional or highly specific design requests.}
  \caption{Failures in generating design images based on participant prompts using DesignFromX and the baseline system. (A) A participant using DesignFromX prompted GenAI to design a pair of glasses with a detachable bridge connected by magnetic pads. (B) A participant using the baseline system prompted GenAI to create a mouse without a scroll wheel. The results illustrate the limitations of GenAI in generating images of unconventional design ideas.}
  \label{fig:userstudy_badexamples}
\end{figure*}

However, we also observed a tradeoff between improving system transparency and promoting the "thinking through" process versus maintaining user control over the GenAI agents. 
Participants reported a slightly lower score for controllability with DesignFromX compared to the baseline system (DesignFromX: M=2.792, SD=0.884; Baseline: M=3.000, SD=1.022; p=0.394, r=-0.228). 
The low controllability of both systems is a result of the limitations of GenAI on image generation. 
We provide one failure example of each system where the GenAI failed to generate the design participants prompted, as shown in Figure \ref{fig:userstudy_badexamples}.
As one participant remarked, \textit{“I like that the system finds the features for me, but I’d prefer to be able to edit and adjust them based on the suggestions” (P5).}
Another participant also pointed out, \textit{"But I think it would be helpful to add some human descriptions to enhance its accuracy. It makes mistakes sometimes. " (P19)}
This was because DesignFromX was developed to explore more designs and did not enable editing on the prompt to tailor the design visualization by GenAI.
The limitations of the image generation capabilities in the GenAI negatively impacted some metrics, resulting in slight improvement but non-significant findings, including collaborative (DesignFromX: M=3.375, SD=0.875; Baseline: M=2.917, SD=0.929; p=0.053, r=0.483) and match goal (DesignFromX: M=3.042, SD=1.042; Baseline: M=2.750, SD=1.073; p=0.271, r=0.284).
Although DesignFromX provided more accurate analysis of design features, enhancing transparency and thoroughness in the human-AI collaborative design process, the constraints of the underlying image generation model restricted overall system performance on these metrics. 
This limitation affected both systems equally, as they utilized the same image generation model.
We further discussed this concern and improvements in Section 8.

\begin{figure}[h]
  \centering
    \includegraphics[width=1\linewidth]{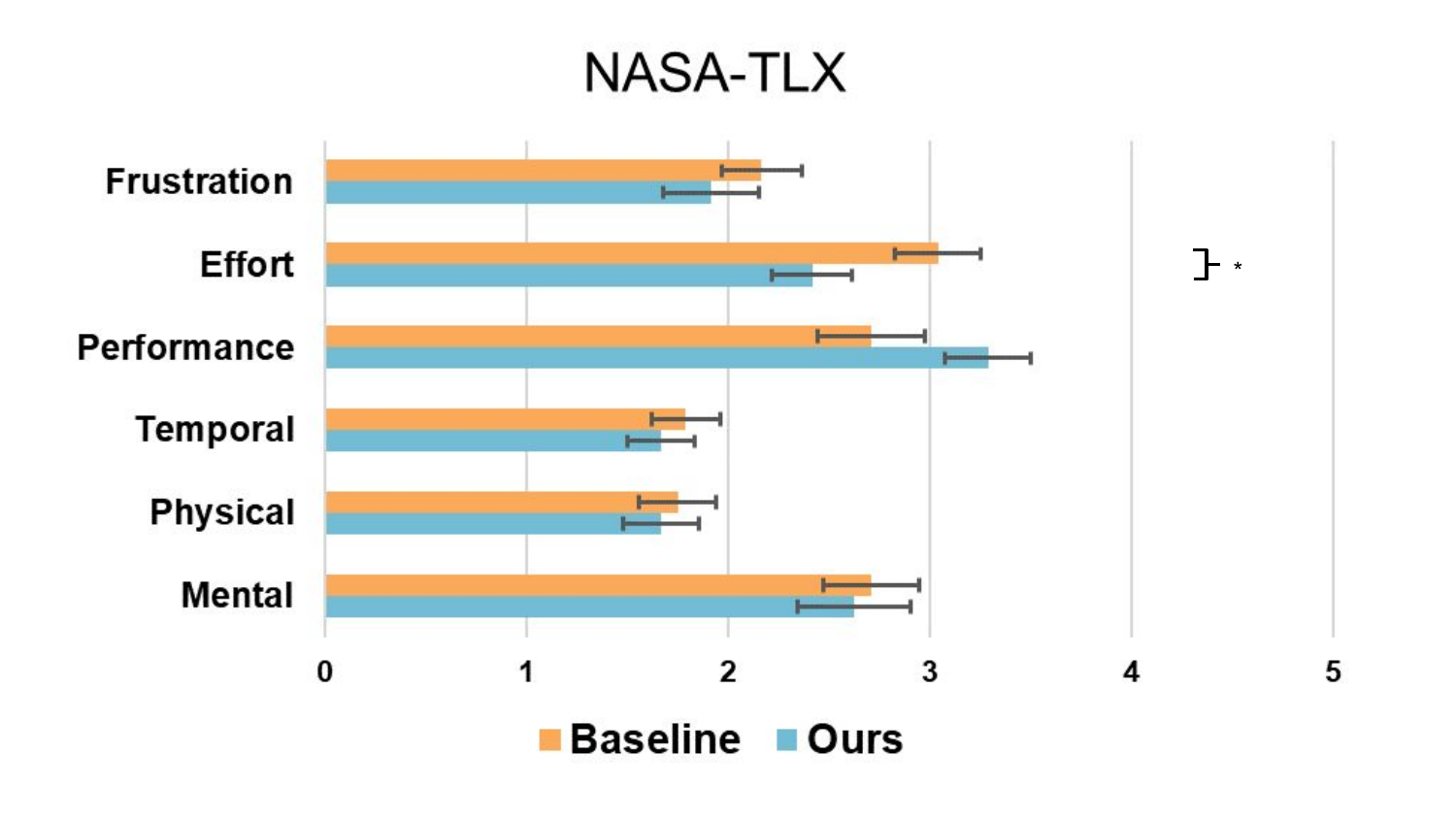}
    \Description[NASA Task Load Index report for DesignFromX.]{This figure shows participants' self-reported workload using the NASA Task Load Index (NASA-TLX), comparing DesignFromX and the baseline system across six dimensions: Frustration, Effort, Performance, Temporal demand, Physical demand, and Mental demand. Participants rated each dimension on a scale from 1 (low) to 5 (high). The results indicated a statistically significant reduction in Effort when using DesignFromX compared to the baseline system, suggesting DesignFromX required less effort from users. No significant differences were found in the other five dimensions.}
    \caption{Statistics of self-reported NASA-TLX using our and baseline systems. * indicates groups with significant differences (p<=0.05).
     The result illustrated significant improvement on the Effort using DesignFromX than the baseline system.}

    \label{fig:NASA-TLX}
\end{figure}

\subsubsection{Engaging in Product Design}

As shown in Figure \ref{fig:self_exp_genai}, participants perceived notably higher levels of enjoyment when using DesignFromX than the baseline system (DesignFromX: M=3.750, SD=0.794; Baseline: M=2.875, SD=0.992; p=0.002, r=0.7). 
The system also significantly improved participants' immersion (DesignFromX: M=3.542, SD=0.588; Baseline: M=2.792, SD=1.021; p=0.003, r=0.635). 

As one participant noted, \textit{"I actually think this system is very interesting. It easily gives people a feeling of wanting to keep trying more design features" (P7). }
These results suggested that the system effectively encouraged consumers to engage more deeply in product design tasks by freeing them on feature analysis, description, and formulating prompts.

Additionally, DesignFromX significantly outperformed the baseline in supporting design exploration (DesignFromX: M=3.917, SD=0.830; Baseline: M=2.417, SD=0.974; p=0.003, r=0.765).
While both systems showed non-significiant results in worth effort (DesignFromX: M=3.25, SD=0.829; Baseline: M=2.958, SD=1.06, p=0.286, r=0.267;) and expressiveness DesignFromX: M=3.292, SD=0.978; Baseline: M=2.792, SD=0.957, p=0.072, r=0.465), our system consistently scored higher across all the evaluated metrics.
This was because of the interactive design experience provided by our system.

Similar to the findings from collaboration with GenAI, the limitations in image generation capabilities restricted participants from fully expressing their ideas within the system and obtaining results that matched their expectations.

\begin{figure}[htp]
  \centering
  \includegraphics[width=1\linewidth]{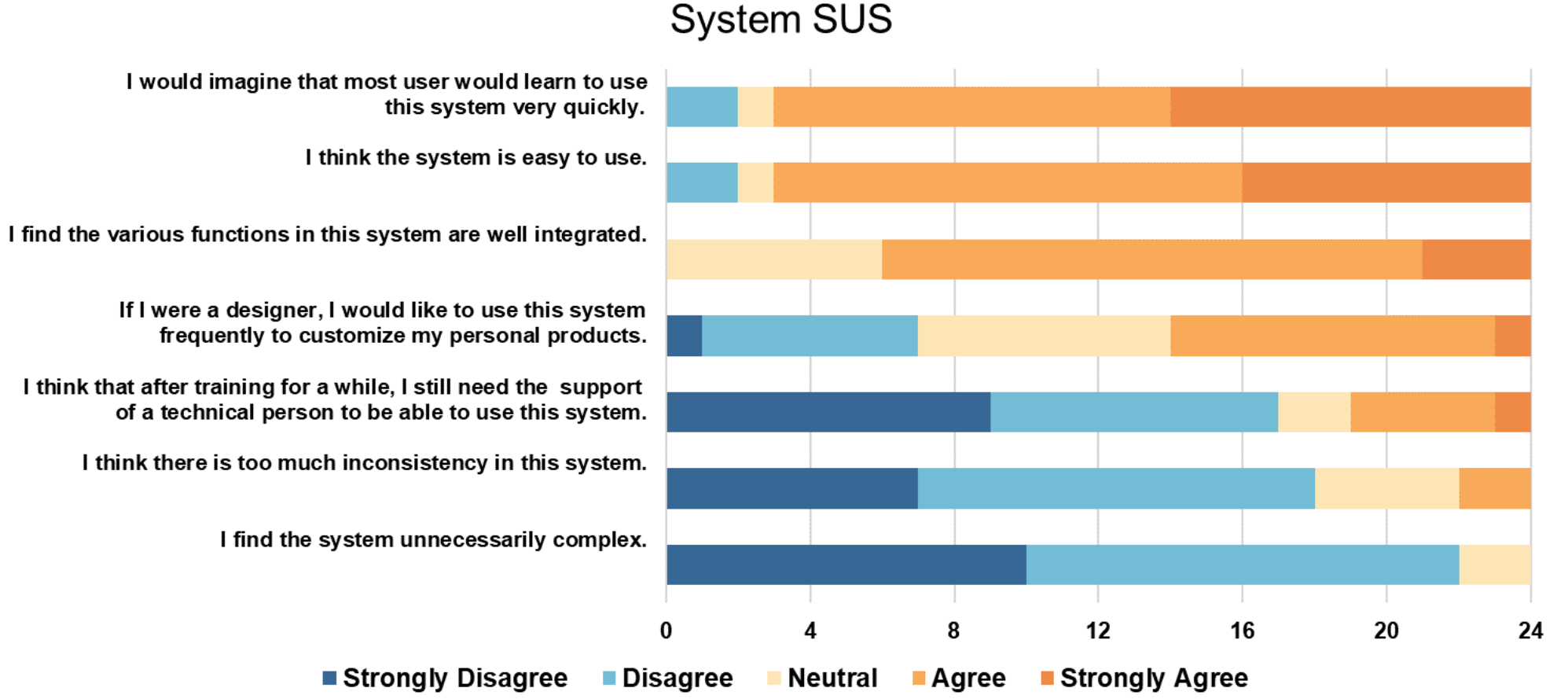}
  \Description[System Usability Scale report for DesignFromX.]{This figure illustrates participants' responses to the System Usability Scale (SUS) survey for DesignFromX. The survey includes statements about the ease of use, learning speed, system complexity, integration of functions, and the potential need for technical support. Participants indicated high agreement that DesignFromX is easy to use, quick to learn, and has well-integrated functions. They generally disagreed with statements suggesting the system is unnecessarily complex, inconsistent, or requires extensive technical support or learning. Overall, results highlight a positive user perception of the usability of DesignFromX.}
  \caption{System usability scores of DesignFromX. Most of the participants consider DesignFromX as easy to learn and use.}
  \label{fig:study2_3 SUS}
\end{figure}
\subsubsection{Overall workload and system usability}

As shown in Figure \ref{fig:NASA-TLX}, DesignFromX significantly reduced the effort needed to complete the design task (DesignFromX: M=3.042, SD=0.974, Baseline: M=2.471, SD=1.042, p=0.037, r=-0.729). 
The participant appreciated the intuitive interactions provided by the system to support design composition.
As one of the participants noted, \textit{"I really like the feature that you can click anywhere on the reference image, then the system automatically provides features that can be used for design." (P11)}
Another participant noted, \textit{"It is very easy to use, especially the brush function that allows you to select what you want to change." (P10)}
For metrics without significant differences—frustration, performance, temporal demand, physical demand, and mental demand—the results suggested that both the baseline system and DesignFromX imposed comparable cognitive and physical workloads on participants. 
The lack of significant differences indicated that the inherent complexity of design tasks influenced workload perceptions similarly across both systems.

The questionnaire on the system usability of DesignFromX collected the subjective feedback from the participants.
The result indicated that the participant appreciated using DesignFromX to support their product design tasks.
The participant reported that DesignFromX was easy to learn, provided consistent performance on the designated task, and featured well-integrated functions.

\subsection{Final Product Design}
\label{sec:final_results}

To answer the research question on the final product design, \textbf{\textit{"RQ3: How do the final products created with DesignFromX differ from those produced using the baseline system?"}}, we collected the expert-reported evaluation and self-reported evaluation on both systems.
The design experts were recruited from among lecturers and teaching assistants of a university-level product design class. 
Each participant's score was calculated as the average rating provided by the three experts.

\subsubsection{Self-Reported Final Design Evaluation}

\begin{figure*}[htp]
  \centering
  \includegraphics[width=\linewidth]{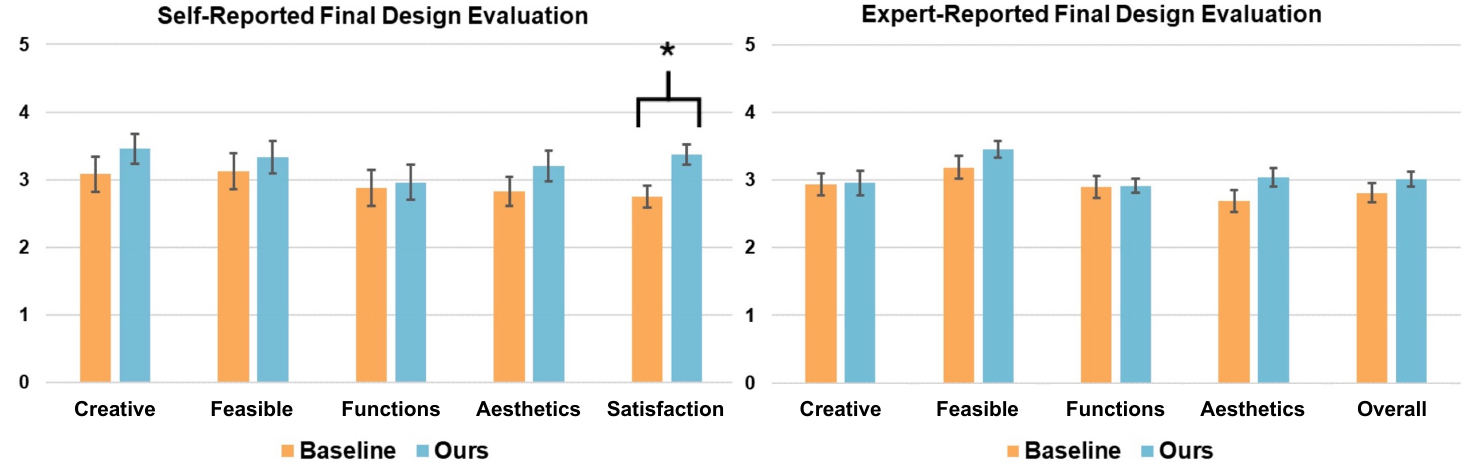}
    \Description[self-report and expert report final design evaluation.]{This figure compares the self-reported and expert-reported evaluations of final designs produced using the DesignFromX system and the baseline system. Participants rated their designs across five dimensions: Creative, Feasible, Functions, Aesthetics, and Satisfaction (Self-reported) or Overall (Expert-reported), on a scale from 1 (low) to 5 (high). In self-reported evaluations, participants using DesignFromX showed significantly higher satisfaction with their final designs compared to the baseline. However, no significant differences emerged in the other 4 dimensions. Expert evaluations did not show significant differences between the two systems across any dimension, although DesignFromX scored slightly higher in feasibility, aesthetics and overall quality. These findings suggest that DesignFromX notably improves participants' personal satisfaction with their designs, despite experts observing no major differences between the systems.}
  \caption{Statistics of self and expert reported final design using our and baseline systems. * indicates groups with significant differences (p<=0.05).
     The result illustrated significant improvement on the self-satisfaction using DesignFromX than the baseline system.}
     \label{fig:self_exp_report_final_concept_env.}
\end{figure*}

Based on the Wilcoxon signed-rank test, the participants perceived significant improvement in their satisfaction with the final design using DesignFromX (DesignFromX: M=3.375, SD=0.711, Baseline: M=2.75, SD=0.794, p=0.013, r=0.639), as shown in Fig. \ref{fig:self_exp_report_final_concept_env.}.
In the post-session interview, over 70\% of the participants (18 out of 24) agreed that they would like to use DesignFromX to design their daily products in the future.
This was because DesignFromX supported designers to explore more designs, giving the participant more options to select based on their own preferences.
Although the participant perceived slight improvements in the other metrics of the final design, the self-reported final design evaluation was not significantly different between DesignFromX and the baseline system.
Additionally, while participants appreciated the creativity and aesthetic aspects of the designs generated with both systems, they were less satisfied with the functionality and feasibility of their designs. 
This was because of the observation that participants, mostly inexperienced designers, expressed uncertainty about the functionality and feasibility of their design.
As one participant noted, \textit{"It seems functional, but I’m not completely sure" (P3)}, while another remarked, \textit{"I don’t know much about manufacturing, but this design doesn’t seem too difficult to build" (P12).}
For the metrics without significant differences—creative, feasible, functions, and aesthetics—the results indicated comparable evaluations between our system and the baseline. 
Participants and experts perceived both systems as similarly effective in supporting creativity, feasibility, functionality, and aesthetic aspects of the final designs. 
The absence of significant differences suggested inherent constraints in the generative AI or similar fundamental performance between the two systems in these specific aspects.

We further discussed this concern in the following Section 8.

\subsubsection{Expert-Reported Final Design Evaluation}

The expert also reported a slight improvement in the overall quality of the final report. 
However, based on the Wilcoxon signed-rank test, there was no significant difference between expert-reported evaluations of the two systems over the five metrics.
Considering that both systems used the same GenAI to develop the designs, it is reasonable that the final design quality between the two systems was not significant.
Also, there was a divergence in opinions between the experts and participants regarding the creativity and feasibility of the final designs. 
While experts viewed the GenAI-supported design as less creative than the participants perceived it to be, they regarded these designs as more feasible to implement than the participants anticipated.
This was because the visualization of participants-created designs was generated by GenAI and followed a similar pattern to existing ones.

\begin{figure*}[htp]
  \centering
  \includegraphics[width=1\linewidth]{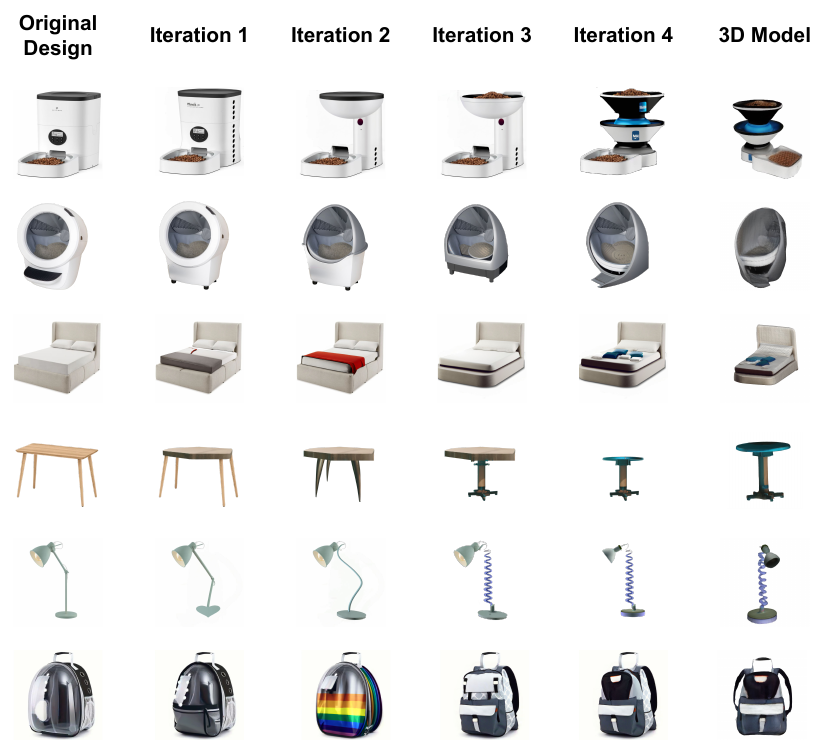}
  \Description[Examples of design iterations using DesignFromX during the user study.]{This figure showcases multiple examples of product design iterations created using DesignFromX during a user study. Each row represents a different product category, including pet feeders, cat litter boxes, beds, tables, lamps, and backpacks. Columns show the progression from an original design through four iterative modifications, culminating in a final 3D model. Across iterations, products gradually incorporate significant changes, such as structural adjustments, added functionalities, varied aesthetics, and personalized features. For example, a pet feeder evolves from a basic cylindrical design to a multi-tiered structure with enhanced functionality; a simple lamp gains a flexible, spiral arm design over iterations. This iterative process illustrates the system's ability to support creative exploration and personalization.}
  \caption{Examples of design iterations using DesignFromX during the user study. The system provides promising results overall. While participants iteratively explored the product design space and used GenAI to modify product images, a decline in image quality was observed, particularly when participants attempted to explore unconventional ideas and incorporate personalized features into the original product.}
     \label{fig:user_iter_results}
\end{figure*}

Additionally, we present examples of iterative design and the final product designs explored during the user study using DesignFromX, as shown in ~\autoref{fig:user_iter_results}. While participants iteratively explored the product design space and used GenAI to modify product images, a decline in image quality was observed, particularly when participants attempted to explore unconventional ideas and incorporate personalized features into the original product.

\section{Discussion and Future Work}

From the system user study, we observed that DesignFromX successfully fulfilled our design goals of supporting consumers in exploring different designs for their everyday products. 
Specifically, participants using DesignFromX composed more design features from references compared to the baseline system (DG1, DG2). 
The system also enhanced the number of generated designs by streamlining feature composition, making the process more structured and efficient (DG3). 
Furthermore, participants reported that DesignFromX offered a more enjoyable and immersive experience than the baseline system (DG4). Ultimately, participants regarded DesignFromX as an accessible and user-friendly tool for exploring the product design space. 
In this section, we discuss the results of the user study and highlight limitations to address in future improvements.

\subsection{Compromise with GenAI: Consumers' Lack of Critical Thinking in Product Design}

Unlike professional designers, who are driven by the pursuit of novelty and feasibility, consumers approach the design process with less critical evaluation. While they are interested in and enjoy the variety of alternative products offered by the system, they often lack deep, structured thinking about how to improve the designs holistically \cite{yang2009observations,song2022design}. 
From the user study, we observed that participants tended to be satisfied with the outputs provided by the system, as long as they noticed changes resulting from their actions.
This characteristic makes consumers more likely to accept compromises in the images generated by GenAI rather than making an effort to steer the generation.

Our observation suggests that consumers prioritize the experience of engaging with the design process over achieving specific design outcomes. 
However, this unintentional and less-structured exploration offers valuable insights for professional designers. 
By analyzing the features that resonate most with consumers, professional designers can uncover patterns of preference and identify key features that are widely appreciated. 
This understanding can be especially beneficial when designing personalized products, enabling professionals to balance user preferences with structural and functional improvements.

To further support consumers in achieving more intentional and critical design exploration, future improvement on DesignFromX could include features such as guided prompts or interactive suggestions that encourage users to consider structural and functional aspects of their designs. This would not only enhance the system’s usability but also help bridge the gap between consumer-driven and professional-level design approaches.

\subsection{Consumers as Determined Designers: Preferring Narrowed Product Features Over Explorative Design}

Unlike professional designers who aim to explore every aspect of a design, consumers tend to act as determined designers, focusing on specific features that align with their personal needs and preferences. 
Our observations indicate that most participants utilized the diverse features provided by the system to initiate their thinking. 
However, they quickly converged on the features that interested them most.

\begin{quote}
\emph{"At first, my ideas were pretty vague, and the system gave me a lot of inspiration. But as my thoughts became clearer and I knew exactly what I wanted to design, having so many options wasn’t as helpful—it didn’t make it easy to add the specific features I was looking for." (P2)}
\end{quote}

Another participant echoed this sentiment, stating: 
\begin{quote}
\emph{"If I know the features I want and can describe them, I’d rather just tell the system to use those features directly." (P21)}
\end{quote}

Prior research on design theory emphasizes the balance between divergent and convergent thinking, where early-stage ideation benefits from broad exploration (divergence), while later stages require a focused approach to refine and develop specific designs \cite{frich2021digital,wadinambiarachchi2024effects,choi2024creativeconnect}.
However, consumers are not tasked with designing a universally perfect product; instead, they aim to make a product perfect for themselves. 
This highlights the strong need to prioritize support for convergent design over the divergence typically emphasized in product design.

Building on these insights, incorporating additional system features that align with this consumer characteristic, such as filtering options or user-driven constraints, could significantly enhance the efficiency of DesignFromX in helping consumers explore and refine their designs.

\subsection{Fight for creativity: Human Creativity Constrained by human-GenAI shared representations}

The advancement in GenAI offers humans greater opportunities to express their ideas by using text-to-image GenAI through descriptive prompts.
This form of ideation, creation, and representation rely on a human-AI shared representation of the creative contents, including images \cite{liu2022opal,chung2023promptpaint}, texts \cite{cummings2024generative}, and graphs \cite{jiang2023graphologue}.
The human-GenAI shared representation encourages the exploration of designs by arguing for human creativity with GenAI's fast visualization ability, but also raised concerns about constraining human creativity \cite{doshi2023generative,li2024value}.

DesignFromX is a design support tool aims at helping consumers explore various products based on reference surrounding in their daily environment.
While it enabled broader explorations by identifying design features from a wide range of other products, we observed a conflict between human creativity and the representations generated by GenAI, as shown in ~\autoref{fig:userstudy_badexamples}.
As a GenAI-powered system, the users of DesignFromX sometimes had to "fight" with the GenAI to achieve their desired outcomes. 
In some cases, human designers found themselves conforming to the AI's suggestions rather than fully exploring their own creative instincts.
One participant described this experience with a metaphor:
\begin{quote}
\emph{"It’s like riding a furious bull that you have to put in so much effort just to steer it in the direction you want to go. Every time you think you’ve got control, it feels like the system pushes back or veers off in its own way, making it hard to stay on track with your original idea. Instead of feeling like a collaborative process, it can sometimes feel like a constant struggle to wrestle it into doing what you need." (P9)}
\end{quote}
This struggle stems from the gap between the human language and the AI model's behavior, as well as between the humans' anticipated design in mind and their language expressed.

From these observations, we envision that the system can be further enhanced by leveraging recent advancements in controllable generative models \cite{zhang2023adding}. These GenAI models emphasize controllability, enabling outputs that more closely align with human intent \cite{bai2022training, wang2023aligning, song2024preference}.

\subsection{Beyond images: Generative Design for Virtual and Physical Prototyping, and Market
Feedback}

DesignFromX employed GenAI to create visual representations of designs, supporting consumers in exploring various designs for everyday products. 
We envision that such generative design tools could play a pivotal role in a broader context of product design by encouraging end-user participation in designing, prototyping, and testing their own product.

By integrating AI-generated designs with virtual prototyping tools \cite{berg2017industry,urban2021designing}, DesignFromX narrowed the gap between product design and the user experience. 
It provided real-time feedback on key design features such as functionality, ergonomics, and aesthetics. 
Additionally, the virtual 3D models created through human-AI collaborative design could be directly utilized in manufacturing workflows \cite{lee2024impact}, further bridging the gap between ideation and production.

Another critical aspect in which DesignFromX extended its capabilities exist in collecting and analyzing early market feedback \cite{jindal2016designed,jin2016understanding}.
By encouraging end-users to explore different design features, professional designers are exposed to more precise consumer insights and market demands, reducing the likelihood of costly redesigns and product failures.

\subsection{DesignFromX for Professional Designers}

Currently, DesignFromX is designed as an accessible, low-barrier design support system that promotes consumer-driven design. 
Its functionalities, including feature analysis and design composition, can further benefit advanced hobbyists and professional designers by facilitating professional exploration of the design space. 
For instance, Large Language Models (LLMs) can analyze multiple products and filter design features based on the specific requirements highlighted by experts. 
Recent studies have explored leveraging LLMs for design space exploration, demonstrating their effectiveness in example organization and management \cite{suh2024luminate,duan2024conceptvis}.

Moreover, the expertise of professional designers allows them to critically evaluate the generated design examples rather than compromise on infeasible results. 
During expert evaluations, as shown in \autoref{fig:self_exp_report_final_concept_env.}, we identified that scores provided by design experts were generally lower compared to self-evaluations. 
Expert designers expressed particular concerns regarding the feasibility of generated examples, assessing them from professional design perspectives, such as constraints and manufacturing capabilities.

Consequently, additional features are required to enhance the current system to better support expert designers. 
For example, assisting expert designers in crafting precise prompts that clearly reflect their professional design intents—rather than vague instructions based on superficial concepts—could substantially improve outcomes. 
Additionally, the generative model must generate images accurately reflecting these precise prompts and conditions, granting designers greater control over the image generation process \cite{zhang2023adding} and providing increased flexibility to leverage their creativity \cite{choi2024creativeconnect,shi2023understanding,duan2025investigating}.

Beyond image generation, a robust professional-oriented system should incorporate modeling and simulation capabilities to seamlessly integrate DesignFromX into the existing design processes, such as supporting parametric CAD modeling \cite{duan2025parametric,gonzalez2023introducing} and manufacturing \cite{gmeiner2023exploring}. 
With ongoing advancements in foundational models and human-computer interaction research \cite{shi2023hci}, we envision the development of professional design support systems that significantly reduce the effort required for designers to conceptualize, prototype, and fabricate real-world products.

\subsection{Limitations and Future work}

Currently, we developed the DesignFromX prototype to study how to support consumers in the product design of everyday products.
While participants provided positive feedback, particularly regarding its ability to help them explore designs, we identified several limitations that highlight areas for future improvement.

\subsubsection{Fine Control Over Generated Designs}

The system currently lacked fine control over the generated visual representations of new designs, such as the ability to tweak shapes or adjust the placement of added features. 
\begin{quote}
\emph{"Sometimes, I want to make more precise adjustments, like dragging the mouse to tweak the curvature of the lamp stand exactly how I imagine it. It’s frustrating when the system only offers preset shapes or automatic changes that don’t quite match what I have in mind. Being able to directly control the details, even something small like a curve, would make the (design) process feel much more intuitive and satisfying." (P7)}
\end{quote}
Introducing such interactive controls \cite{dang2022ganslider} can enable users to further customize the generated designs to better match their preferences

\subsubsection{Selection of Granular Design Features}

DesignFromX recommended design features by providing text-based descriptions. 
While these descriptions ensured comprehensive explanations of the features, some participants desired to select granular features for their designs.
\begin{quote}
\emph{"I think there could be a function that allows you to select more specific traits from the recommended ones. Right now, its descriptions are very professional and detailed, but sometimes they don’t exactly match what I need." (P9)}
\end{quote}

This insight indicates a future improvement in providing more flexible options to use the recommended features.

\subsubsection{Tailoring Prompts for Convergent Design Composition}

To allow consumers to focus on exploring more designs, DesignFromX did not enable them to tailor the prompts for generating visual representations of the designs.
To address this limitation, future improvements should focus on allowing users to refine and customize prompts, ensuring that the visual representations better reflect their design intents.

\subsubsection{Enhancing GenAI's Understanding of Design Language}

Finally, participants using both DesignFromX and the baseline system noted that the current GenAI sometimes struggled to understand specific design language, such as descriptions of styles and mechanisms. 
As one participant remarked:
\begin{quote}
\emph{"I noticed that it would use some repetitive descriptions, like 'easy to use.' It (GenAI) didn’t seem to fully understand or describe the specific features in context" (P14). }
\end{quote}

We envision that the development of design-oriented GenAI will lead to a better understanding of product design languages. These advancements can provide a more accurate representation of the products, thus enabling the professional designer to leverage the power of GenAI in their workflows.

\subsubsection{Integrating with More Fundamental Models}

As discussed in the previous subsection, the current system can be enhanced by integrating it with more foundational models to better support professional design workflows. 
For instance, the generated 3D models could be further parameterized to integrate with parametric modeling software, enabling advanced optimization and streamlined fabrication processes. 
Additionally, the current feature analysis could leverage these 3D models—instead of relying solely on images—to perform precise constraint-based evaluations, such as verifying whether new structures meet dimensional or stress-related constraints within the overall product design.

\section{Conclusion}

In conclusion, DesignFromX demonstrated its potential to support consumers in exploring and composing product designs by leveraging GenAI's ability to explore useful design features from other products. 
We achieved this by conducting a formative study with novice designers to understand the barriers in their product design practices.
Then, we derived design goals to develop DesignFromX, a consumer-driven design support tool that aimed at empowering consumers to explore the designs of their daily products.
We conducted a two-phase user study to evaluate the effectiveness of DesignFromX in fulfilling its design goals.
Through our user studies, we found that the system provided participants with a more engaging and enjoyable design experience on their daily products.

DesignFromX sets a foundation for further advancements in GenAI-supported product design, with the goal of empowering end-users to seamlessly navigate through the design space and generate meaningful, personalized design outcomes.

\begin{acks} 

We wish to thank all reviewers for their invaluable feedback. This work is partially supported by the NSF under the Future of Work at the Human-Technology Frontier (FW-HTF) 1839971. The authors also acknowledge the Feddersen Distinguished Professorship Funds and a gift from Thomas J. Malott. Any opinions, findings, and conclusions expressed in this material are those of the authors and do not necessarily reflect the views of the funding agency.

\end{acks}
\bibliographystyle{ACM-Reference-Format}
\bibliography{ref}

\appendix
\section{Appendix}
\subsection{Evaluation Result on the Quality of System Modules}
We present the evaluation result on the quality of system modules.

The testing data revealed that the component query module achieved an accuracy of 0.90 in recognizing the labels of segmented components, with an Intersection Over Union (IoU) score of 0.839 when compared to the human-labeled ground truth. These results demonstrate the module's capability to accurately identify the product components selected by users.

\begin{table}[htp]
\centering

\begin{tabular}{c|cc}
\specialhline{0.42pt}
\hline
\specialhline{0.4pt}
 & IOU & Label Accuracy \\
\hline
Mean & {0.839} & 0.90 \\
Std & {0.105} & 0.30 \\

\specialhline{0.42pt}
\hline
\specialhline{0.4pt}
\end{tabular}
\caption{Expert evaluation of region detection and labeling performance.}
  \Description[Expert evaluation on component segmentation and labeling function.]{This table summarizes the expert evaluation results for region detection and labeling performance, measured by two metrics: Intersection over Union (IOU) and Label Accuracy. The average IOU score is high (0.839), indicating accurate region detection, with a standard deviation of 0.105. The label accuracy average is 0.90, indicating high correctness in labeling, with a standard deviation of 0.30.}
\end{table}

The results indicated that the feature analysis module integrated into the DesignFromX delivered high-quality descriptions for all design features, achieving an average accuracy and overall score exceeding 4 out of 5.
These findings highlight the module's capability to effectively support users in articulating design features from existing products.

\begin{table*}[htp]
\centering
\begin{tabular}{c|c|c|cccc}

\specialhline{0.42pt}
\hline
\specialhline{0.4pt}
\multicolumn{2}{c}{}&& Accuracy & Fluency & Relevancy& Satisfaction \\
\hline
\multirow{8}{*}{Aesthetic}
    &\multirow{2}{*}{Color}&Mean&4.667 & 4.643 & 4.667 & 4.690 \\
    & &Std&0.713 & 0.610 & 0.678 & 0.556 \\
    \cline{2-7}
    &\multirow{2}{*}{Style}&Mean&4.571 & 4.405 & 4.571 & 4.595 \\
    & &Std&0.583 &0.726 & 0.583 & 0.580 \\

    \cline{2-7}
    &\multirow{2}{*}{Texture}&Mean&4.405 & 4.381 & 4.405 & 4.405 \\
    & &Std&0.847 & 0.815 & 0.818 & 0.818 \\

    \cline{2-7}
    &\multirow{2}{*}{Shape}&Mean&4.262 & 4.262 & 4.357 & 4.310 \\
    & &Std&0.818 & 0.726 &0.781 & 0.831 \\

\hline
\multirow{8}{*}{Functional} 
    &\multirow{2}{*}{Structure}&Mean&4.238 & 4.357 & 4.190 & 4.214 \\
    & &Std&0.811 & 0.684 & 0.823 & 0.832 \\

    \cline{2-7}
    &\multirow{2}{*}{Ergonomic}&Mean&3.762 & 3.857 & 3.643 & 3.738 \\
    & &Std&1.191 & 1.059 & 1.151 & 1.196 \\

    \cline{2-7}
    &\multirow{2}{*}{Mechanism}&Mean&4.119 & 4.214 & 4.262 & 4.190 \\
    & &Std&1.095 & 1.059 & 1.001 & 1.052 \\

    \cline{2-7}
    &\multirow{2}{*}{Electronic}&Mean&4.333 & 4.357 & 4.405 & 4.405 \\
    & &Std& 1.127 & 1.042 & 1.048 & 1.025 \\

\specialhline{0.42pt}
\hline
\specialhline{0.4pt}
\end{tabular}
\label{tab: feature identification}
\caption{Expert evaluation of feature identification performance.}
  \Description[Expert evaluation on Accuracy, Fluency, Relevancy, and Satisfaction for feature analysis.]{This table summarizes expert evaluations on feature identification performance across aesthetic features (color, style, texture, shape) and functional features (structure, ergonomic, mechanism, electronic). Each feature was rated on four criteria: accuracy, fluency, relevancy, and satisfaction, on a scale presumably from 1 (lowest) to 5 (highest). Aesthetic Features generally received high ratings, particularly for color (average ~4.7) and style (~4.6), indicating strong performance across all four criteria. Among those functional features,  structure and electronic received consistently high scores (around 4.3), reflecting strong accuracy, fluency, relevancy, and satisfaction in identification performance. Overall, the evaluations suggest the system performed well on aesthetic and functional feature identification.}
\end{table*}

\subsection{Formative Study Interview Questions}

We summarized the interview question (IQs) for the formative study, including:

\textit{IQ1:"In your daily life, have you ever thought about modifying an existing product or make a new product? Could you describe what product it is and why you would like to make this modification?"}

\textit{IQ2:"When designing a product, what do you consider the most important factor and why?"}

\textit{IQ3:"Have you encountered any difficulties or frustrations while designing the products in this experiment? "}

\textit{IQ4:"Could you describe your experience using AI-generated design examples? What advantages did you notice and what challenges did you encounter?"}

\textit{IQ5:"If you were provided with an AI tool to assist in designing everyday products, what specific features or capabilities would you expect from it?"}

\subsection{Data Report for Overall Worklaod and Final Porduct Evaluation}

We report the detailed data for the non-significant metrices in overall workload evaluation, including, frustration (DesignFromX: M=1.917, SD=1.176; Baseline: M=2.167, SD=0.963; p=0.265, r=-0.271), performance (DesignFromX: M=3.292, SD=1.042; Baseline: M=2.708, SD=1.301; p=0.084, r=0.397;), temporal demand (DesignFromX: M=1.667, SD=0.816; Baseline: M=1.792, SD=0.833; p=0.477, r=-0.214), physical demand (DesignFromX: M=1.667, SD=0.917; Baseline: M=1.75, SD= 0.944; p=0.658, r=-0.123), and mental demand (DesignFromX: M=2.625, SD=1.377; Baseline: M=2.708, SD=0.977; p=0.688, r=-0.116)

We also report the detailed data for the non-significant metrices in final product evaluation, including self-reported final design evaluation: creative (DesignFromX: M=3.458, SD=1.103; Baseline: M=3.083, SD=1.283; p=0.306, r=0.235), feasible (DesignFromX: M=3.333, SD=1.204; Baseline: M=3.125, SD=1.296; p=0.61, r=0.12), functions (DesignFromX: M=2.958, SD=1.268; Baseline: M=2.875, SD=1.296; p=0.84, r=0.042), aesthetics (DesignFromX: M=3.208, SD=1.103; Baseline: M=2.833, SD=1.049; p=0.202, r=0.301), expert-reported final design evaluation: creative (DesignFromX: M=2.958, SD=0.871; Baseline: M=2.938, SD=0.785; p=0.877, r=0.037), feasible (DesignFromX: M=3.458, SD= 0.606; Baseline: M=3.188, SD=0.845; p=0.367, r=0.207), functions (DesignFromX: M=2.917, SD=0.482; Baseline: M=2.896, SD=0.794; p=0.825, r=0.052), aesthetics (DesignFromX: M=3.042, SD=0.674; Baseline: M=2.688, SD=0.778; p=0.104, r=0.394), overall quality (DesignFromX: M=3.021, SD=0.541; Baseline: M=2.812, SD=0.689; p=0.267, r=0.297).


\end{document}